\documentclass[aps,prd,floatfix,twocolumn,showpacs,amsmath,amssymb,nofootinbib]{revtex4-1}
\usepackage{graphicx}
\usepackage{color}
\usepackage{times}
\usepackage{inputenc}
\usepackage{bm}
\usepackage{ulem}
\usepackage{multirow}
\usepackage{float}
\usepackage{url}
\usepackage{natbib}
\usepackage{mathrsfs}
\usepackage{physics}
\usepackage{comment}
\usepackage[table,xcdraw]{xcolor}
\usepackage{booktabs}
\usepackage{comment}
\usepackage{soul}
\usepackage[caption=false]{subfig}
\usepackage[colorlinks=true,citecolor=blue,urlcolor=blue,linkcolor=blue]{hyperref}
\usepackage{tikz,xcolor,hyperref}
\definecolor{lime}{HTML}{A6CE39}
\DeclareRobustCommand{\orcidicon}{%
	\begin{tikzpicture}
	\draw[lime,fill=lime] (0,0) 
	circle [radius=0.16] 
	node[white] {{\fontfamily{qag}\selectfont \tiny ID}};
	\draw[white,fill=white] (-0.0625,0.095) 
	circle [radius=0.007];
	\end{tikzpicture}
	\hspace{-2mm}
}
\foreach \x in {A,...,Z}{%
	\expandafter\xdef\csname orcid\x\endcsname{\noexpand\href{https://orcid.org/\csname orcidauthor\x\endcsname}{\noexpand\orcidicon}}
}

\begin{document}
\title{Trace anomaly and interior curvature of neutron stars in energy-momentum squared gravity}
\author{Ratikanta Swain\orcidD{}}
\email{swainratikanta2000@gmail.com}
\author{Sayantan Ghosh\orcidA{}}
\email{sayantanghosh1999@gmail.com}
\author{Bharat Kumar\orcidB{}}
\email{kumarbh@nitrkl.ac.in}

\affiliation{Department of Physics and Astronomy, National Institute of Technology, Rourkela 769008, India}
\date{\today}
\begin{abstract}
In energy-momentum squared gravity (EMSG), the spacetime inside a neutron star is sourced by effective thermodynamic variables that need not coincide with the physical fluid pressure and energy density. It is therefore an open question whether the trace anomaly of dense matter---the QCD measure of how strongly conformal symmetry is broken---still organizes interior profiles and curvature in the same way it does in general relativity (GR). We adopt a clear matter--geometry separation: the trace anomaly is computed from the fluid sector alone, while spacetime curvature scalars are built from the variables that actually source the modified Tolman-Oppenheimer-Volkoff equations. For five relativistic mean-field equations of state, the radial trace-anomaly profiles increase monotonically from core to surface in all accepted EMSG models, as in GR, but split systematically with the EMSG coupling strength; the splitting grows with stellar compactness. Despite this deformation, curvature invariants still fall onto organized bands when plotted against the trace anomaly, extending the GR thermodynamic-geometric correspondence. The Ricci contraction shows the tightest organization, whereas the Ricci scalar remains the most equation-of-state sensitive. EMSG effects are modest for observationally accessible stars but largest in stiff, ultracompact configurations, indicating that the trace anomaly remains a useful thermodynamic label for interior geometry even when gravity couples nonlinearly to matter.
\end{abstract}
\maketitle

\section{Introduction}
\label{intro}
Neutron stars (NSs) are unique laboratories for matter under extreme conditions\cite{Glendenning,shapiro,Chandrasekhar_1964,Rezzolla_2018}. Formed in core-collapse supernovae\cite{supernovae,Chanmugam_1977,Rezzolla_2002}, they compress matter to several times nuclear saturation density in cores only a few tens of kilometers across\cite{Lattimer2015,Psaltis2008,Li:2011vx,Miller_2019,Riley_2021}. Their hydrostatic equilibrium is governed by the Tolman--Oppenheimer--Volkoff (TOV) equations\cite{OppenheimerVolkoff,Wald:1984rg}, supplemented by an equation of state (EOS) that encodes QCD matter, nuclear interactions, and possible exotic phases\cite{rajagopal2001condensed,alford2008color,APR,Rezzolla_2018,Bharat_and_Landry}. In this sense, NSs probe both the cold EOS of dense matter and general relativity (GR) far beyond solar-system tests\cite{Einstein1915,Will2014,Clifton_2012,Kopeikin2014}.

The observational landscape has transformed dramatically over the past decade. Radio timing of millisecond pulsars, including massive objects such as PSR~J0740$+$6620 and PSR~J0952$-$0607, has established that NSs can reach masses above $2\,M_{\odot}$\cite{Miller_2019,Riley_2021,Miller_2021,Crombie2022,GW190814}, placing strong lower bounds on the maximum mass supported by the EOS. X-ray pulse-profile modeling with NICER has begun to constrain mass--radius relations for individual pulsars\cite{Miller_2019,Riley_2019,Miller_2021,Riley_2021}, while the gravitational-wave event GW170817\cite{Abbott_2017,GW170817,doi:10.1126/science.aap9811,Abbott_2020} provided the first direct access to tidal deformability and hence to the compactness of canonical-mass stars\cite{PhysRevLett.121.091102,PhysRevX.9.011001,Fragione_2021,Annala_2020}. Follow-up analyses combining nuclear experiments, chiral effective field theory, and astrophysical data have narrowed the space of viable EOS models\cite{Annala_2020,Jiang_2020,VQCD1,VQCD2,Xiao}. At the same time, precision tests of GR with binary pulsars\cite{PhysRevX.11.041050,Psaltis2008} and the emerging possibility of measuring NS moment of inertia through pulsar timing\cite{Kent_yagi_2013,Kent_yagi_2015,Haskell_2015,Gupta_2018} have renewed interest in whether NS phenomenology can discriminate among gravitational theories\cite{Yagi:2013bca,CAPOZZIELLO2011167,OLMO,OLMO20201,lobo,Sotiriou,CQ}.

Interpreting such rich data is complicated by the fact that most interior quantities depend sensitively on the EOS. Standard macroscopic parameters---maximum mass, radius, tidal deformability, and moment of inertia---are correlated through the celebrated I--Love--$\mathcal{Q}$ relations\cite{Yagi:2013bca,Kent_yagi_2013,Kent_yagi_2015,Haskell_2013,Haskell_2014,Haskell_2015}, yet they still require an EOS choice to translate into local thermodynamic information. A complementary approach is to identify dimensionless interior variables that encode the stiffness of matter directly and, where possible, correlate with geometry or global observables with reduced model dependence. The trace anomaly and spacetime curvature invariants have emerged as particularly promising candidates in this context\cite{ren2026tracingtraceanomalydense,TraceCurvature,ILoveCurvature2025,Cai_2026,Cai2024compactness,PhysRevLett.129.252702}. At the same time, several classes of modified gravity theories predict departures from GR precisely in regions where matter is present, making NS interiors natural targets for strong-field tests of extended frameworks such as $f(R)$ gravity\cite{NOJIRI201159,doiNOJIRI}, $f(R,T)$ theories\cite{Harko2011,PhysRevD.84.024020,PhysRevD.91.044034}, and energy-momentum squared gravity (EMSG)\cite{PhysRevD.94.044002,PhysRevD.98.024031,Nari,Akarsu,EMSG_OAkarsu,EMSG_NAlam,Cipriano_2024,Ghosh2026}.

\subsection{Trace anomaly and curvature in General Relativity}
The trace anomaly $\Delta$ [Eq.~\eqref{eq:Delta}] measures how far cold dense matter departs from the conformal symmetry expected of massless QCD\cite{PhysRevLett.129.252702,ren2026tracingtraceanomalydense,Cai_2026}. Running couplings and finite quark masses break this symmetry at the densities reached in NS cores, so perturbative QCD predicts that the anomaly vanishes only at asymptotically high density\cite{QCD1,QCD2,QCD3,QCD4,QCD5,Lin_2004}. Fujimoto \textit{et al.}\cite{PhysRevLett.129.252702} proposed the normalized trace anomaly as a dimensionless measure of conformality and showed that the speed of sound can be decomposed into contributions tied to $\Delta$, making the latter more informative about the approach to the conformal limit than the instantaneous value of $c_s^2$ alone. Thermodynamic stability and causality restrict $\Delta$ to the range $-2/3\leq\Delta<1/3$ in NS matter\cite{PhysRevLett.129.252702,ren2026tracingtraceanomalydense}, and three recent GR developments motivate the present study.

First, Cai \textit{et al.}\cite{Cai_2026,Cai2024compactness} showed from the intrinsic structure of the TOV equations that the pressure-to-energy ratio decreases monotonically outward from the stellar center for any viable static solution, independently of the input EOS. The trace anomaly therefore attains its minimum in the core and increases toward the dilute surface; in the compactness-scaling approach of Cai \textit{et al.}\cite{Cai2024compactness}, this central thermodynamic state can be inferred model-insensitively from mass--radius or redshift measurements. Second, Ren and Lin\cite{ren2026tracingtraceanomalydense} constructed quasi-universal relations linking the radial profile of the pressure-to-energy ratio $X\equiv P/\mathcal{E}$---and hence of $\Delta$ through $\Delta=1/3-X$---to compactness, normalized moment of inertia, and tidal deformability, using 45 observationally viable EOS models. Their fitted surfaces reproduce the interior profile band to within about $10\%$ EOS scatter and were applied to NICER targets and to the GW170817 tidal constraint, yielding a central trace anomaly of $0.1770^{+0.0365}_{-0.0432}$ for a $1.4\,M_\odot$ star. Third, Garibay \textit{et al.}\cite{TraceCurvature} examined an ensemble of viable EOSs and found that curvature invariants correlate systematically with $\Delta$: for about $50\%$ of the models, negative Ricci curvature appears somewhere in the interior, predominantly for stiff and compact stars; the central Ricci scalar vanishes at a zero-curvature compactness $\mathscr{C}_0=0.26911\pm0.00004$, where $\Delta_c=0$, and the ensemble reaches $\Delta_{\min}=-0.227$ near the maximum-mass limit. Danarianto \textit{et al.}\cite{ILoveCurvature2025} extended the I--Love--$\mathcal{Q}$ program by linking tidal deformability and moment of inertia to local, surface, and volume-averaged curvature scalars in GR. Together, these works suggest that $\Delta$ acts as a reduced thermodynamic coordinate for both the matter sector and the interior geometry of NSs in GR\cite{TraceCurvature,ren2026tracingtraceanomalydense,Yagi:2013bca,Kent_yagi_2013}.

\subsection{The EMSG complication}
Energy-Momentum Squared Gravity (EMSG) modifies GR by coupling the spacetime curvature to the square of the matter stress--energy\cite{PhysRevD.94.044002,PhysRevD.98.024031,AkarsuNew,AKARSU2023101305,EMSG_OAkarsu,EMSG_NAlam}. The theory coincides with GR in vacuum but introduces density-dependent corrections inside matter, where the coupling constant $\alpha$ is constrained to be extremely small by cosmological and binary-pulsar data\cite{EMSG_OAkarsu,Nazari1,PhysRevD.93.023501} yet can alter NS structure when central densities are high\cite{EMSG_NAlam,Pinku-PRD_2023,Pinku_jcap_2023,Pinku_mnras_2023}. For a perfect fluid the field equations can be recast in GR-like form with effective pressure and energy density [Eqs.~\eqref{eq:Eeff}--\eqref{eq:Peff}] that differ from the physical fluid variables.

Existing EMSG studies have focused on modified mass--radius relations, tidal deformability, $f$-mode frequencies, and binary-pulsar bounds\cite{Nari,PhysRevD.98.024031,EMSG_NAlam,Ghosh2025,Harish-fmode_2022,Zhao_gmode_2022,Nazari,Board2017}. Most recently, Ghosh \textit{et al.}\cite{Ghosh2026} computed the Kretschmann, Ricci, and Weyl scalars in EMSG as functions of baryon density and radius, showing that positive (negative) $\alpha$ enhances (suppresses) curvature and that hadron--quark phase transitions can leave distinct imprints in the curvature profiles. That work established how effective sourcing reshapes the interior curvature structure, but it did not ask whether the \textit{fluid-sector} trace anomaly $\Delta$, defined exactly as in GR, still correlates with curvature when the two sectors are treated separately. Nor did it test the trace-anomaly--observable organization of Ren and Lin\cite{ren2026tracingtraceanomalydense} or the curvature--$\Delta$ correspondence of Garibay \textit{et al.}\cite{TraceCurvature} beyond $\alpha=0$. The gap is conceptual as well as numerical: if one replaces the fluid variables by the effective ones in the definition of $\Delta$, the thermodynamic label itself changes and direct comparison with the GR literature is lost. The central question addressed here is therefore the following.
\medskip
\noindent\textit{When gravity is sourced by effective variables in EMSG, does the trace anomaly $\Delta$, defined only from the fluid sector, still provide an organizing parameter for interior spacetime curvature?}
\medskip

\subsection{Strategy and main results of this work}
Our approach is deliberately asymmetric and follows the matter--geometry separation introduced in Sec.~\ref{sec:matter-geometry}: $\Delta$ is computed from the fluid sector alone, whereas the curvature invariants $\mathcal{K}$, $\mathcal{W}$, $\mathcal{R}$, and $\mathcal{F}$ are built from the effective variables that source the modified TOV equations. Any correlation between curvature and $\Delta$ in EMSG is therefore nontrivial, because the horizontal axis carries a theory-independent thermodynamic label while the vertical axis reflects modified effective sourcing [Eqs.~\eqref{eq:separation}--\eqref{eq:IofDelta}].

Solving the modified TOV system for the five relativistic mean-field EOSs of Sec.~\ref{sec:eos} (NL3, IOPB-I, G3, SINPA, GM1), we find:
\begin{enumerate}
    \item In all accepted EMSG integrations, the radial profile of $\Delta$ increases monotonically outward and remains EOS structured, numerically mirroring the outward-increasing GR trend of Cai \textit{et al.}\cite{Cai_2026} without invoking their GR-only proof, but acquires an $\alpha$-dependent splitting that grows with compactness and is strongest for smaller tidal deformability and smaller normalized moment of inertia.
    \item Curvature invariants collapse onto organized one-parameter bands when plotted against $\Delta$, extending the GR curvature--$\Delta$ correspondence of Garibay \textit{et al.}\cite{TraceCurvature} and complementing the I--Love--curvature relations of Danarianto \textit{et al.}\cite{ILoveCurvature2025} in a modified-gravity setting.
    \item EMSG corrections are largest for ultracompact configurations, yet remain smooth and regular over the coupling interval of Eq.~\eqref{eq:alpha-range}.
    \item At fixed compactness, tidal deformability, and moment of inertia, the $\alpha$-induced splitting of $\Delta(r)$ remains modest for observationally accessible parameters, while the Ricci contraction shows the tightest organization among the four invariants.
\end{enumerate}
These findings show that nonlinear gravity deforms, but does not erase, the thermodynamic--geometric organization already established in GR. The remainder of the paper is organized as follows. Section~\ref{TF} presents the EMSG framework, curvature invariants, the matter--geometry separation, tidal deformability, moment of inertia, and the EOS models. Section~\ref{RnD} reports the numerical results. Section~\ref{sec:limitations} discusses scope and limitations. Section~\ref{SnC} summarizes the implications. We use metric signature $(-,+,+,+)$ and geometrized units $G=c=1$ unless SI curvature units are explicitly displayed.
\section{Theoretical Framework}
\label{TF}
\subsection{Energy-Momentum Squared Gravity (EMSG)}
EMSG belongs to the broader class of theories in which gravity is coupled nonlinearly to the energy--momentum tensor\cite{Harko2010,Harko2011,PhysRevD.94.044002,Katirci2014,PhysRevD.96.123517,PhysRevD.97.024011,Faraoni,PhysRevD.109.104055}. Such theories are often equivalent, under field redefinitions, to GR with nonminimal matter coupling\cite{AkarsuNew,Board2017,PhysRevD.101.064021,PhysRevD.101.124006}. In the standard formulation adopted here, the Einstein--Hilbert action is modified by a term quadratic in $T_{\mu\nu}T^{\mu\nu}$. The full action reads\cite{AkarsuNew,PhysRevD.94.044002}:
\begin{equation}
    S = \int \left[ \frac{1}{2\kappa} R + f\left(\mathcal{L}_m,\,
    g_{\mu\nu}T^{\mu\nu},\,T_{\mu\nu}T^{\mu\nu}\right) 
    + \mathcal{L}_m \right] \sqrt{-g}\,d^4x,
    \label{eq:action}
\end{equation}
where $R$ is the Ricci scalar, $\kappa = 8\pi G$, $\mathcal{L}_m$ is the matter Lagrangian, $T_{\mu\nu}$ is the energy--momentum tensor, and $f$ is a function of matter scalars. For a perfect fluid, $T_{\mu\nu}$ follows from the matter Lagrangian\cite{Katirci2014,PhysRevD.94.044002,PhysRevD.97.024011,PhysRevD.96.123517,EMSG_OAkarsu,EMSG_NAlam}:
\begin{equation}
    T_{\mu\nu} = -\frac{2}{\sqrt{-g}} 
    \frac{\delta(\sqrt{-g}\,\mathcal{L}_m)}{\delta g^{\mu\nu}}
    = g_{\mu\nu}\mathcal{L}_m - 2\frac{\partial \mathcal{L}_m}
    {\partial g^{\mu\nu}},
    \label{eq:EMT}
\end{equation}
which depends strongly on the metric tensor. The total matter Lagrangian is\cite{AkarsuNew}
\begin{equation}
    \mathcal{L}_m^{\mathrm{tot}} = \mathcal{L}_m + f,
    \label{eq:Ltot}
\end{equation}
which leads to a modified energy--momentum tensor
\begin{equation}
\begin{aligned}
    T_{\mu\nu}^{\mathrm{tot}} & = 
    -\frac{2}{\sqrt{-g}} 
    \frac{\delta(\sqrt{-g}\,\mathcal{L}_m^{\mathrm{tot}})}
    {\delta g^{\mu\nu}} \\
    & = -\frac{2}{\sqrt{-g}} 
    \frac{\delta(\sqrt{-g}\,\mathcal{L}_m)}{\delta g^{\mu\nu}}
    -\frac{2}{\sqrt{-g}} 
    \frac{\delta(\sqrt{-g}\,f)}{\delta g^{\mu\nu}}.
    \label{eq:Ttot}
\end{aligned}
\end{equation}
Thus,
\begin{equation}
     T_{\mu\nu}^{\mathrm{tot}} = T_{\mu\nu} + T_{\mu\nu}^{\mathrm{mod}},
    \label{eq:Ttot2}
\end{equation}
where the modification may be written as\cite{AkarsuNew}
\begin{equation}
     T_{\mu\nu}^{\mathrm{mod}} =-\frac{2}{\sqrt{-g}} 
    \frac{\delta(\sqrt{-g}\,f)}{\delta g^{\mu\nu}}
    = f\,g_{\mu\nu} 
    - 2f_{T^2}\,\theta_{\mu\nu},
    \label{eq:Tmod2}
\end{equation}
with
\begin{equation}
        f_{T^2} = \frac{\partial f}{\partial(T_{\rho\sigma}T^{\rho\sigma})},
    \qquad
    \theta_{\mu\nu} = \frac{\delta(T_{\rho\sigma}T^{\rho\sigma})}
    {\delta g^{\mu\nu}}.
    \label{eq:fT2}
\end{equation}
We adopt the standard EMSG choice
\begin{equation}
    f = \alpha\,T_{\mu\nu}T^{\mu\nu},
    \label{eq:EMSG_choice}
\end{equation}
where $\alpha$ is the coupling parameter, giving\cite{Ghosh2025,Katirci2014,PhysRevD.94.044002,PhysRevD.97.024011,PhysRevD.96.123517,EMSG_OAkarsu,EMSG_NAlam}
\begin{equation}
    T_{\mu\nu}^{\mathrm{mod}} = \alpha\,T_{\rho\sigma}T^{\rho\sigma}\,
    g_{\mu\nu} - 2\alpha\,\theta_{\mu\nu}.
    \label{eq:Tmod3}
\end{equation}

The field equations then read
\begin{equation}
    G_{\mu\nu} = \kappa T_{\mu\nu} + \kappa T_{\mu\nu}^{\mathrm{mod}},
    \label{eq:field1}
\end{equation}
or, explicitly,
\begin{equation}
G_{\mu\nu} = \kappa T_{\mu\nu} + \kappa\alpha 
    \left( g_{\mu\nu} T_{\sigma\epsilon}T^{\sigma\epsilon} 
    - 2\theta_{\mu\nu} \right).
\end{equation}
For a perfect fluid\cite{Faraoni,PhysRevD.109.104055},
\begin{equation}
    T_{\mu\nu} = (\mathcal{E} + P)\,u_\mu u_\nu + P\,g_{\mu\nu},
    \label{eq:perfectfluid}
\end{equation}
with $\mathcal{E}$ the energy density, $P$ the pressure, and $u^\mu$ the four-velocity satisfying $u^\mu u_\mu = -1$.

Substituting Eqs.~\eqref{eq:perfectfluid}--\eqref{eq:Tmod3} into Eq.~\eqref{eq:field1} and simplifying, one obtains
\begin{align}
    G_{\mu\nu} &= \kappa \mathcal{E}\left[\left(1 + \frac{P}{\mathcal{E}}\right)
    u_\mu u_\nu + \frac{P}{\mathcal{E}}\,g_{\mu\nu}\right] \notag \\
    &+ \kappa\alpha \mathcal{E}^2\left[2\left(1 + \frac{4P}{\mathcal{E}} 
    + \frac{3P^2}{\mathcal{E}^2}\right)u_\mu u_\nu 
    + \left(1 + \frac{3P^2}{\mathcal{E}^2}\right)g_{\mu\nu}\right].
    \label{eq:Gmunu}
\end{align}

Equation~\eqref{eq:Gmunu} can be recast in GR-like form,
\begin{equation}
    G_{\mu\nu} = \kappa T_{\mu\nu}^{\mathrm{eff}},
    \label{eq:Geff}
\end{equation}
with
\begin{equation}
    T_{\mu\nu}^{\mathrm{eff}} = 
    (\mathcal{E}_{\mathrm{eff}} + P_{\mathrm{eff}})\,u_\mu u_\nu 
    + P_{\mathrm{eff}}\,g_{\mu\nu}.
\end{equation}
This defines the effective thermodynamic variables
\begin{equation}
    \mathcal{E}_{\mathrm{eff}} = \mathcal{E} + \alpha \mathcal{E}^2 
    \left(1 + \frac{8P}{\mathcal{E}} + 3\frac{P^2}{\mathcal{E}^2}\right),
    \label{eq:Eeff}
\end{equation}
\begin{equation}
    P_{\mathrm{eff}} = P + \alpha \mathcal{E}^2 
    \left(1 + 3\frac{P^2}{\mathcal{E}^2}\right).
    \label{eq:Peff}
\end{equation}

\subsection{Modified Tolman--Oppenheimer--Volkoff Equations in
EMSG}

We consider a static, spherically symmetric line element\cite{Wald:1984rg,Schwarzschild:1916uq}:
\begin{equation}
    ds^2 = -e^{2\nu(r)}dt^2 + e^{2\lambda(r)}dr^2 
    + r^2\left(d\theta^2 + \sin^2\theta\,d\phi^2\right),
    \label{eq:metric}
\end{equation}
where $\nu(r)$ and $\lambda(r)$ are the metric functions. Solving the modified Einstein equations in the EMSG formalism yields the modified TOV system:

\begin{equation}
    \frac{dm}{dr} = 4\pi r^2 \mathcal{E}\left[1 + \alpha \mathcal{E} 
    \left(1 + \frac{8P}{\mathcal{E}} + 3\frac{P^2}{\mathcal{E}^2}\right)\right],
    \label{eq:TOV1}
\end{equation}
\begin{align}
    \frac{dP}{dr} &= -\frac{m\mathcal{E}}{r^2}
    \left(1 + \frac{P}{\mathcal{E}}\right)
    \left(1 - \frac{2m}{r}\right)^{-1} \notag \\
    &\times \left[1 + \frac{4\pi r^3 P}{m} 
    + \frac{4\pi r^3 \alpha \mathcal{E}^2}{m}
    \left(1 + 3\frac{P^2}{\mathcal{E}^2}\right)\right] \notag \\
    &\times \left[1 + 2\alpha \mathcal{E}\left(1 + \frac{3P}{\mathcal{E}}\right)\right]
    \left[1 + 2\alpha \mathcal{E}\left(c_s^{-2} 
    + \frac{3P}{\mathcal{E}}\right)\right]^{-1}.
    \label{eq:TOV2}
\end{align}
Equations~\eqref{eq:TOV1}--\eqref{eq:TOV2} generalize the standard Tolman--Oppenheimer--Volkoff system\cite{Wald:1984rg,Schwarzschild:1916uq,OppenheimerVolkoff}. They are integrated from $r=0$ with $m(0)=0$ and central pressure $P(0)=P_c$ until $P(R)=0$, which defines the stellar radius $R$ and gravitational mass $M=m(R)$. We verify that $c_s^2\equiv dP/d\mathcal{E}\leq 1$ and $dP/d\mathcal{E}>0$ at all interior points for every accepted model\cite{shapiro,PhysRevLett.129.252702}.
\subsection{Spacetime curvature invariants}
\label{sec:curv-defs}
For the static, spherically symmetric line element of Eq.~\eqref{eq:metric}, the Riemann tensor yields coordinate-independent scalars that diagnose the local geometry\cite{Curvature1,Curvature2,Curvature3,Ghosh2026,Carroll,Hawking:1973uf,shakerin,Breu_2016,TraceCurvature}. We employ the Kretschmann scalar $\mathcal{K}\equiv\sqrt{R_{\mu\nu\rho\sigma}R^{\mu\nu\rho\sigma}}$, the Weyl scalar $\mathcal{W}\equiv\sqrt{C_{\mu\nu\rho\sigma}C^{\mu\nu\rho\sigma}}$, the Ricci scalar $\mathcal{R}\equiv g^{\mu\nu}R_{\mu\nu}$, and the Ricci contraction $\mathcal{F}\equiv\sqrt{R_{\mu\nu}R^{\mu\nu}}$, following the conventions of Garibay \textit{et al.}\cite{TraceCurvature} and Ghosh \textit{et al.}\cite{Ghosh2026}. In GR these quantities are fully determined by $(\mathcal{E},P)$ through the field equations\cite{Wald:1984rg,Schwarzschild:1916uq}. In EMSG they are evaluated from $(\mathcal{E}_{\mathrm{eff}},P_{\mathrm{eff}})$ because the metric responds to the effective source.

All curvature quantities in what follows are computed from $\mathcal{E}_{\mathrm{eff}}$ and $P_{\mathrm{eff}}$, reflecting the nonlinear response of spacetime to matter\cite{Ghosh2026,EMSG_NAlam,Pinku_mnras_2023}. The correction terms scale as $\alpha\mathcal{E}^{2}$ and therefore grow most rapidly in the dense core, where even cosmologically small $\alpha$ can produce order-unity shifts in the invariants at fixed $\rho_B$\cite{EMSG_NAlam,Pinku-PRD_2023,Pinku_mnras_2023}.

\subsubsection{Kretschmann Scalar}
The Kretschmann scalar $\mathcal{K}(r)$ measures the total spacetime curvature, including contributions from both the local effective matter source and the macroscopic gravitational field\cite{Curvature1,Curvature2,Ghosh2026,TraceCurvature}. In Eq.~\eqref{eq:Kretschmann}, the terms quadratic in $(\mathcal{E}_{\mathrm{eff}},P_{\mathrm{eff}})$ encode local sourcing, whereas the pieces involving $m(r)$ describe the accumulated gravitational potential. Their interplay sets the radial falloff of $\mathcal{K}$ from the dense core to the surface.

\begin{align}
    \mathcal{K}(r) &= \left[(8\pi)^2\left(3\mathcal{E}_{\mathrm{eff}}^2 
    + 3P_{\mathrm{eff}}^2 + 2\mathcal{E}_{\mathrm{eff}}P_{\mathrm{eff}}\right) 
    \right. \notag \\
    &\left. - \frac{128\pi \mathcal{E}_{\mathrm{eff}}\,m(r)}{r^3} 
    + \frac{48\,m^2(r)}{r^6}\right]^{1/2}.
    \label{eq:Kretschmann}
\end{align}

\subsubsection{Weyl Scalar}
The Weyl scalar $\mathcal{W}(r)$ measures the tidal component of the gravitational field and vanishes at the center for regular spherical solutions\cite{Curvature3,ILoveCurvature2025}. As a function of radius it grows toward the outer layers, where the enclosed mass distribution departs most strongly from the local energy density; as a function of baryon density it is subdominant in the core (Sec.~\ref{sec:curv-density}).

\begin{equation}
    \mathcal{W}(r) = \left[\frac{4}{3}
    \left(\frac{6m(r)}{r^3} - 8\pi \mathcal{E}_{\mathrm{eff}}\right)^2
    \right]^{1/2}.
    \label{eq:Weyl}
\end{equation}

\subsubsection{Ricci Scalar}
The Ricci scalar $\mathcal{R}(r)$ is tied directly to the trace of the effective energy--momentum tensor through Eq.~\eqref{eq:Ricci} and therefore reflects the local balance between $\mathcal{E}_{\mathrm{eff}}$ and $P_{\mathrm{eff}}$\cite{TraceCurvature,Carroll}. Because $\mathcal{R}\propto(\mathcal{E}_{\mathrm{eff}}-3P_{\mathrm{eff}})$, it is the most sensitive invariant to sign changes in the effective pressure--energy ratio and can become negative in stiff, compact cores\cite{TraceCurvature}.

\begin{equation}
    \mathcal{R}(r) = \kappa\left(\mathcal{E}_{\mathrm{eff}} - 3P_{\mathrm{eff}}\right).
    \label{eq:Ricci}
\end{equation}

\subsubsection{Ricci Contraction}
The Ricci contraction $\mathcal{F}(r)$ provides a positive-definite measure of matter-induced curvature that depends quadratically on $(\mathcal{E}_{\mathrm{eff}},P_{\mathrm{eff}})$ without the cancellations present in $\mathcal{R}$\cite{TraceCurvature,ILoveCurvature2025}. In GR it corresponds to Garibay \textit{et al.}'s\cite{TraceCurvature} invariant $\mathcal{J}=\sqrt{R_{\mu\nu}R^{\mu\nu}}$, and both Garibay \textit{et al.}\cite{TraceCurvature} and Danarianto \textit{et al.}\cite{ILoveCurvature2025} report strong correlations between such contractions and global observables; we therefore treat $\mathcal{F}(\Delta)$ as a principal diagnostic in Sec.~\ref{sec:curv-delta}.

\begin{equation}
    \mathcal{F}(r) = \sqrt{64\pi^2 \left(\mathcal{E}_{\mathrm{eff}}^2 + 3P_{\mathrm{eff}}^2\right)}.
    \label{eq:F}
\end{equation}

\subsection{Trace Anomaly}

The trace anomaly is defined exclusively from the fluid variables $(\mathcal{E},P)$\cite{ren2026tracingtraceanomalydense,Cai2024compactness}. It provides a dimensionless measure of the departure from conformal symmetry\cite{TraceCurvature,Trace4,Trace5,Trace6,PhysRevLett.129.252702} and links microscopic stiffness to macroscopic observables\cite{Lin_2004,Wei_2020,Xiao,Zhao_2022}. Because $\Delta$ depends only on $\phi\equiv P/\mathcal{E}$, it is insensitive to an overall scaling of the EOS and instead reflects the local stiffness of matter. Along a static NS profile, the radial variation of $\Delta$ traces the transition from a strongly interacting core to a dilute envelope\cite{Lattimer2015,PhysRevC.103.035810,Bikram_2023}.
\begin{equation}
    \Delta(r) = \frac{\mathcal{E}(r) - 3P(r)}{3\mathcal{E}(r)} 
    = \frac{1}{3} - \frac{P(r)}{\mathcal{E}(r)},
    \label{eq:Delta}
\end{equation}


\subsection{Matter--geometry separation}
\label{sec:matter-geometry}
This subsection states the conceptual framework that distinguishes the present analysis from prior work. In GR, Garibay \textit{et al.}\cite{TraceCurvature} correlated curvature invariants with $\Delta$ because both sectors are sourced by the same $(\mathcal{E},P)$. Ren and Lin\cite{ren2026tracingtraceanomalydense} linked the profile $\Delta(r)$ to global observables without invoking modified gravity. Ghosh \textit{et al.}\cite{Ghosh2026} computed EMSG curvature profiles as functions of baryon density and radius, but did not hold the GR definition of $\Delta$ fixed while testing whether geometry still tracks it. Here we combine these threads under one protocol.
In EMSG the metric is sourced by $T_{\mu\nu}^{\mathrm{eff}}$ [Eq.~\eqref{eq:Geff}], not by the fluid tensor $T_{\mu\nu}$ of Eq.~\eqref{eq:perfectfluid}. The effective variables $(\mathcal{E}_{\mathrm{eff}},P_{\mathrm{eff}})$ in Eqs.~\eqref{eq:Eeff}--\eqref{eq:Peff} encode how nonlinear gravity responds to matter, whereas $(\mathcal{E},P)$ continue to describe the thermodynamic state of the stellar fluid itself. Treating these two sets interchangeably would mix modified-gravity effects into the definition of $\Delta$ and obscure the central question posed in Sec.~\ref{intro}.

We therefore adopt an explicit two-sector protocol:
\begin{itemize}
    \item \textbf{Matter sector (theory independent).} The trace anomaly is evaluated only from the fluid variables $(\mathcal{E},P)$ through Eq.~\eqref{eq:Delta}. Because $\Delta$ depends solely on the ratio $P/\mathcal{E}$, it carries the same interpretation in GR and EMSG: it measures how far the local fluid departs from conformal symmetry. No $\alpha$-dependent correction enters its definition.
    \item \textbf{Geometry sector (theory dependent).} Curvature invariants are constructed from the metric generated by $(\mathcal{E}_{\mathrm{eff}},P_{\mathrm{eff}})$ through Eqs.~\eqref{eq:Kretschmann}--\eqref{eq:F}. Symbolically,
    \begin{equation}
        \mathcal{I}(r) = \mathcal{G}\!\left(\mathcal{E}_{\mathrm{eff}}(r),\,P_{\mathrm{eff}}(r)\right),
        \qquad
        \mathcal{I}\in\{\mathcal{K},\mathcal{W},\mathcal{R},\mathcal{F}\}.
        \label{eq:separation}
    \end{equation}
    In EMSG this sector is genuinely modified: $\mathcal{E}_{\mathrm{eff}}\neq\mathcal{E}$ and $P_{\mathrm{eff}}\neq P$ whenever $\alpha\neq 0$.
\end{itemize}
The separation is not merely notational. A quantity such as $\mathcal{R}=\kappa(\mathcal{E}_{\mathrm{eff}}-3P_{\mathrm{eff}})$ [Eq.~\eqref{eq:Ricci}] resembles a trace-anomaly structure, but it is built from effective variables and therefore measures curvature sourced by $T_{\mu\nu}^{\mathrm{eff}}$, not the trace anomaly of Eq.~\eqref{eq:Delta}. Likewise, replacing $(\mathcal{E},P)$ by $(\mathcal{E}_{\mathrm{eff}},P_{\mathrm{eff}})$ in the definition of $\Delta$ would redefine the thermodynamic label itself and prevent a direct comparison with GR studies\cite{ren2026tracingtraceanomalydense,Cai_2026,TraceCurvature}.

Collecting both sectors, each curvature invariant is a composite function of the trace anomaly, the coupling $\alpha$, and the EOS:
\begin{equation}
    \mathcal{I} = f(\Delta;\,\alpha,\,\mathrm{EOS}),
    \label{eq:IofDelta}
\end{equation}
because Eqs.~\eqref{eq:Eeff}--\eqref{eq:Peff} make $(\mathcal{E}_{\mathrm{eff}},P_{\mathrm{eff}})$ nonlinear functions of $(\mathcal{E},P)$ at fixed $\alpha$. In GR, $\mathcal{E}_{\mathrm{eff}}=\mathcal{E}$ and $P_{\mathrm{eff}}=P$, so organized $\mathcal{I}(\Delta)$ bands are expected from prior work\cite{TraceCurvature}. In EMSG the correspondence is \textit{not} guaranteed \textit{a priori}: effective sourcing can in principle decouple geometry from the thermodynamic state even when $\Delta$ remains well defined. The numerical program of Sec.~\ref{RnD} is therefore a direct test of whether Eq.~\eqref{eq:IofDelta} remains sufficiently tight, across five RMF EOSs and the coupling range quoted in Sec.~\ref{sec:numerics}, for $\Delta$ to serve as a reduced coordinate for interior spacetime geometry.

\subsection{Tidal deformability, moment of inertia, and global observables}
\label{sec:tidal}
To compare $\Delta(r)$ with quantities accessible to gravitational-wave and timing observations, we require the tidal deformability $\Lambda$, the compactness $C$, and the normalized moment of inertia $\bar{I}\equiv I/M^3$. In the static-tide limit relevant to binary inspirals, a neutron star subject to an external quadrupolar tidal field $\mathcal{E}_{ij}$ develops an induced mass quadrupole
\begin{equation}
    Q_{ij} = -\lambda\,\mathcal{E}_{ij},
    \label{eq:quadrupole}
\end{equation}
where $\lambda$ is the tidal deformability\cite{Yagi:2013bca,Kent_yagi_2013}. For a nonrotating star, the linearized GR tidal equations of Yagi and Yunes\cite{Yagi:2013bca,Kent_yagi_2013,Kent_yagi_2015} reduce to ordinary differential equations coupled to the background stellar profile. In the present work we apply these GR formulae to the $(\mathcal{E},P)$ background obtained from the modified TOV integration; a fully self-consistent EMSG tidal theory would additionally modify the perturbation sector (Sec.~\ref{sec:limitations}). Introducing perturbation functions $H(r)$ and $\beta(r)\equiv dH/dr$, and writing $f(r)\equiv 1-2m(r)/r$, one integrates the background TOV system together with
\begin{align}
    \frac{dH}{dr} &= \beta, &
    \frac{d\beta}{dr} &= \frac{2}{f}\,H\,\mathcal{A}_t
    + \frac{2}{rf}\,\beta\,\mathcal{B}_t,
    \label{eq:tidal-pert}
\end{align}
where
\begin{align}
    \mathcal{A}_t &= -2\pi\!\left[5\mathcal{E}+9P+\frac{d\mathcal{E}}{dP}(\mathcal{E}+P)\right]
    + \frac{3}{r^2}
    + \frac{2}{f}\!\left(\frac{m}{r^2}+4\pi r P\right)^{\!2},
    \label{eq:At}\\
    \mathcal{B}_t &= -1 + \frac{m}{r} + 2\pi r^2(\mathcal{E}-P),
    \label{eq:Bt}
\end{align}
imposing regular central behavior $H\propto r^2$ and $\beta\propto 2r$ as $r\to 0$ and matching at the stellar surface $P(R)=0$\cite{Yagi:2013bca,Kent_yagi_2013}. The second Love number $k_2$ is obtained from the compactness $C\equiv M/R$ and the surface ratio
\begin{equation}
    y_s \equiv \left.\frac{r\,\beta}{H}\right|_{r=R},
    \label{eq:ysurf}
\end{equation}
through the standard exterior-potential matching formula\cite{Yagi:2013bca,Kent_yagi_2013,Kent_yagi_2015}. The dimensionless tidal deformability used throughout this work is
\begin{equation}
    \Lambda \equiv \frac{\lambda}{M^5} = \frac{2}{3}\,\frac{k_2}{C^5},
    \label{eq:Lambda}
\end{equation}
which is the combination inferred from binary neutron-star mergers such as GW170817\cite{PhysRevLett.121.091102,GW170817,Annala_2020}. Because $\Lambda$ is primarily sensitive to the stiffness of the outer layers, it provides a complementary global constraint on $\Delta(r)$ to compactness and moment of inertia within the I--Love--$\mathcal{Q}$ framework\cite{Yagi:2013bca,Kent_yagi_2013,Kent_yagi_2015,Haskell_2015,ILoveCurvature2025}.

The moment of inertia probes the interior mass distribution and is accessible, in principle, to double-pulsar timing\cite{Yagi:2013bca,Kent_yagi_2013,Kent_yagi_2015,Haskell_2015,Gupta_2018}. In the slow-rotation limit, Hartle--Thorne perturbation theory introduces a frame-dragging function $\omega(r)$\cite{Yagi:2013bca,Kent_yagi_2013,Kent_yagi_2015}. As in the tidal sector, we apply the GR formulae to the $(\mathcal{E},P)$ background obtained from the modified TOV integration (Sec.~\ref{sec:limitations}). Coupled to Eqs.~\eqref{eq:TOV1}--\eqref{eq:TOV2}, the first-order system reads
\begin{align}
    \frac{d\omega}{dr} &= \omega', &
    \frac{d\omega'}{dr} &= -\mathcal{A}_\omega\,\omega' + \mathcal{B}_\omega\,\omega,
    \label{eq:moi-pert}
\end{align}
where
\begin{align}
    \mathcal{A}_\omega &= \frac{4}{r} - \frac{4\pi r(\mathcal{E}+P)}{f},
    \label{eq:Aomega}\\
    \mathcal{B}_\omega &= \frac{16\pi(\mathcal{E}+P)}{f},
    \label{eq:Bomega}
\end{align}
and $f(r)\equiv 1-2m(r)/r$ as above. Regular solutions impose $\omega(0)=1$ and $\omega'(0)=0$, and the integration is terminated at the stellar surface $P(R)=0$\cite{Yagi:2013bca,Kent_yagi_2013}. The moment of inertia follows from the surface values as
\begin{equation}
    I = \frac{R^{4}}{6}\,\frac{\omega'(R)}{\omega(R)},
    \label{eq:I}
\end{equation}
and the dimensionless quantity used in the $\bar{I}$-selected sequences is
\begin{equation}
    \bar{I} \equiv \frac{I}{M^{3}}.
    \label{eq:Ibar}
\end{equation}
Because $\bar{I}$ weights the entire radial mass profile rather than only the outer layers, it is only weakly correlated with $C$ and $\Lambda$ in the I--Love--$\mathcal{Q}$ relations and therefore provides an independent global lever on $\Delta(r)$\cite{Yagi:2013bca,Kent_yagi_2013,Kent_yagi_2015,Gupta_2018,ILoveCurvature2025}.

For the $\Lambda$- and $\bar{I}$-selected sequences analyzed in Sec.~\ref{sec:TA-observables}, the target observables are first mapped to central conditions using GR reference integrations of Eqs.~\eqref{eq:tidal-pert}--\eqref{eq:Lambda} and Eqs.~\eqref{eq:moi-pert}--\eqref{eq:Ibar}, following the sequence construction of Ren and Lin\cite{ren2026tracingtraceanomalydense}. The EMSG profiles $\Delta(r)$, curvature invariants, and compactness are then evaluated by solving Eqs.~\eqref{eq:TOV1}--\eqref{eq:TOV2} at the corresponding central pressure for each $(\mathrm{EOS},\alpha)$.

\subsection{Equation of state models}
\label{sec:eos}
We employ five relativistic mean-field (RMF) EOSs spanning soft to stiff behavior: NL3\cite{NL3EOS,NL3}, IOPB-I\cite{IOPB-I,IOPB-IEOS}, G3\cite{G3EOS,G3}, SINPA\cite{SINPAEOS}, and GM1\cite{GM1EOS}. These models are widely used in NS structure studies\cite{Bharat_and_Landry,Zhao_2022,Tianqi_2022,PhysRevC.103.035810,gc2024} and match the hadronic set employed in recent EMSG curvature analyses\cite{Ghosh2026}. They cover the stiffness range over which Garibay \textit{et al.}\cite{TraceCurvature} reported the strongest $\mathcal{R}$--$\Delta$ correlations in GR.

\subsection{Numerical implementation}
\label{sec:numerics}
For each EOS we integrate Eqs.~\eqref{eq:TOV1}--\eqref{eq:TOV2} from $r=0$ with $m(0)=0$ and central pressure $P_c$ until $P(R)=0$, which defines the stellar radius $R$ and gravitational mass $M=m(R)$. The coupling parameter is varied in the range
\begin{equation}
    -6.07\times 10^{6}\,\mathrm{m}^{2} \leq \alpha \leq 6.07\times 10^{6}\,\mathrm{m}^{2},
    \label{eq:alpha-range}
\end{equation}
bracketing the values used in prior EMSG neutron-star analyses\cite{Ghosh2026,Ghosh2025,EMSG_NAlam,Pinku-PRD_2023}. For each accepted $(\mathrm{EOS},\alpha,P_c)$ solution we evaluate $\Delta$ from $(\mathcal{E},P)$ and the curvature invariants from $(\mathcal{E}_{\mathrm{eff}},P_{\mathrm{eff}})$ using Eqs.~\eqref{eq:Kretschmann}--\eqref{eq:F}, and verify that $c_s^2\equiv dP/d\mathcal{E}$ remains positive and causal throughout the interior\cite{shapiro,PhysRevLett.129.252702}. Compactness follows as $C=M/R$. For the $(C,\Lambda,\bar{I})$ sequences of Sec.~\ref{sec:TA-observables}, the target observables are mapped to central conditions with GR reference integrations following Ren and Lin\cite{ren2026tracingtraceanomalydense}; the EMSG profiles are then computed at the corresponding $(P_c,\alpha)$. GR ($\alpha=0$) baselines are generated with the same procedure for direct comparison.

\section{Results and Discussion}
\label{RnD}
We organize the numerical results around the central question of Sec.~\ref{intro}. Section~\ref{sec:curv-density} establishes how EMSG reshapes the curvature invariants as functions of baryon density, extending the density-based analysis of Ghosh \textit{et al.}\cite{Ghosh2026} but preparing the $\mathcal{I}(\Delta)$ test of Sec.~\ref{sec:curv-delta}. Section~\ref{sec:curv-radial} examines the radial profiles and compares them with the GR I--Love--curvature trends of Danarianto \textit{et al.}\cite{ILoveCurvature2025}. Sections~\ref{sec:TA-profiles} and~\ref{sec:TA-observables} study the fluid-sector trace anomaly $\Delta(r)$ and its response to $\alpha$ at fixed $(C,\Lambda,\bar{I})$, directly confronting the quasi-universal relations of Ren and Lin\cite{ren2026tracingtraceanomalydense} and the monotonicity theorems of Cai \textit{et al.}\cite{Cai_2026}. Section~\ref{sec:curv-delta} performs the principal test: whether Garibay \textit{et al.}'s\cite{TraceCurvature} GR correspondence between curvature and $\Delta$ survives when geometry is sourced by $(\mathcal{E}_{\mathrm{eff}},P_{\mathrm{eff}})$. Finally, Sec.~\ref{sec:obs-sequences} places the EMSG bands in the observational context of NICER and GW170817.

\subsection{Curvature scalars as functions of baryon density}
\label{sec:curv-density}
\begin{figure*}
    \centering
    \includegraphics[width=\textwidth]{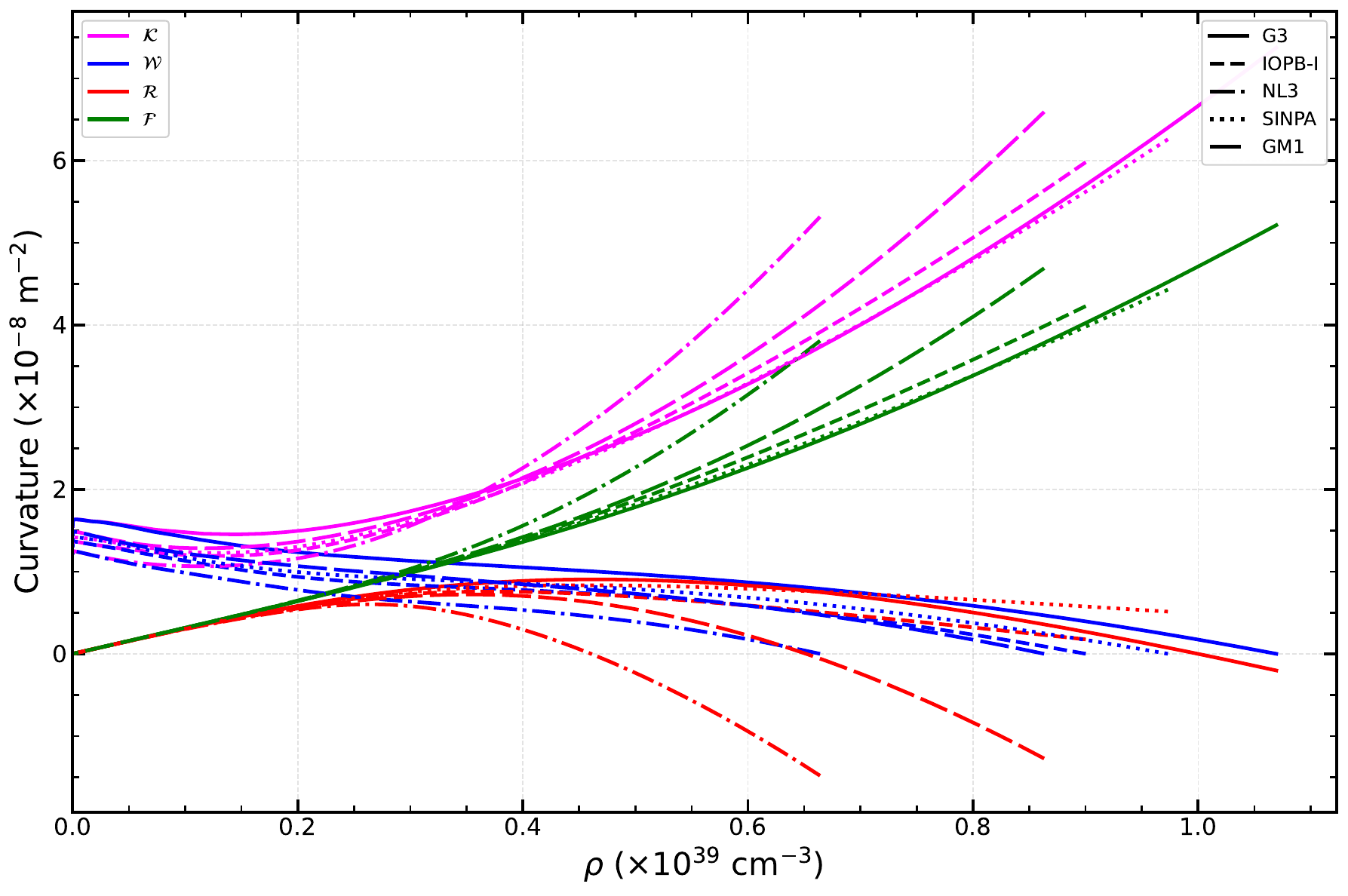}
    \caption{Curvature invariants in EMSG neutron stars as functions of baryon number density $\rho_B$.
    The Kretschmann scalar $\mathcal{K}$, Weyl scalar $\mathcal{W}$, Ricci scalar $\mathcal{R}$, and Ricci contraction $\mathcal{F}$ are evaluated from $(\mathcal{E}_{\mathrm{eff}},P_{\mathrm{eff}})$ for NL3, IOPB-I, G3, SINPA, and GM1.
    Curvature is plotted in units of $10^{-8}\,\mathrm{m}^{-2}$; $\rho_B$ is in units of $10^{39}\,\mathrm{cm}^{-3}$ (nuclear saturation density corresponds to $\rho_B\simeq 0.16$ on this scale). Dashed segments indicate the GR limit $\alpha=0$ where applicable.}
    \label{AllcurvWithBaryonDensity}
\end{figure*}

Figure~\ref{AllcurvWithBaryonDensity} shows the curvature invariants as functions of baryon density $\rho_B$ for the five RMF EOSs at representative values of $\alpha$. In GR, Garibay \textit{et al.}\cite{TraceCurvature} reported similar scalars for an ensemble of observationally viable models and emphasized the link between negative $\mathcal{R}$ and stiff, compact stars. Ghosh \textit{et al.}\cite{Ghosh2026} carried the density-based analysis into EMSG, highlighting sensitivity to core stiffness and to hadron--quark transitions. Our figure follows that EMSG convention---invariants are built from $(\mathcal{E}_{\mathrm{eff}},P_{\mathrm{eff}})$---but is organized to connect directly with the $\mathcal{I}(\Delta)$ correspondence tested in Sec.~\ref{sec:curv-delta}, where the horizontal axis is the fluid-sector label $\Delta$.
Because the EMSG corrections in Eqs.~\eqref{eq:Eeff}--\eqref{eq:Peff} scale as $\alpha\mathcal{E}^2$, the density dependence is not a simple vertical rescaling of the GR curves. At $\rho_B\gtrsim 0.3\times 10^{39}\,\mathrm{cm}^{-3}$ (approximately $2\rho_0$ on the figure scale) the effective source can compete with the linear fluid term even for cosmologically small $\alpha$, steepening the growth of $\mathcal{K}$ and $\mathcal{F}$ with density\cite{EMSG_NAlam,Pinku-PRD_2023,Ghosh2026}. The Kretschmann scalar $\mathcal{K}$ measures the full Riemann contraction and therefore probes the local strength of spacetime curvature, including both matter and gravitational-potential contributions\cite{TraceCurvature,ILoveCurvature2025,Hawking:1973uf}. Its monotonic growth with $\rho_B$ reflects the accumulation of effective energy density in the core, where the metric is most strongly bent by $T_{\mu\nu}^{\mathrm{eff}}$. For the stiffest EOSs we find central values $\mathcal{K} \sim 7 \times 10^{-8}\,\mathrm{m}^{-2}$ and $\mathcal{F} \sim 5 \times 10^{-8}\,\mathrm{m}^{-2}$, consistent in order of magnitude with the GR ensemble of Garibay \textit{et al.}\cite{TraceCurvature}, although EMSG vertically shifts the curves through $(\mathcal{E}_{\mathrm{eff}},P_{\mathrm{eff}})$. Stiff models such as NL3 and GM1 reach the largest curvatures because their large symmetry-energy slope forces $P/\mathcal{E}$ above the conformal value in the supranuclear core\cite{NL3EOS,GM1EOS,PhysRevC.103.035810}, whereas softer models such as IOPB-I and SINPA maintain lower effective sourcing and therefore lower $\mathcal{K}$ and $\mathcal{F}$ at fixed $\rho_B$\cite{IOPB-IEOS,SINPAEOS}.

The Ricci contraction $\mathcal{F}=\sqrt{64\pi^2(\mathcal{E}_{\mathrm{eff}}^2+3P_{\mathrm{eff}}^2)}$ remains positive definite and provides a monotonic measure of how strongly matter sources curvature, without the cancellations present in $\mathcal{R}$. The Weyl scalar behaves differently: it decreases with increasing $\rho_B$ and remains subdominant to $\mathcal{K}$ over most of the interior, because $\mathcal{W}\propto[(6m/r^3)-8\pi\mathcal{E}_{\mathrm{eff}}]^2$ measures the tidal mismatch between enclosed mass and local energy density rather than the local trace sector\cite{Curvature1,Curvature2,ILoveCurvature2025}. At high baryon density the Ricci-sector terms dominate and tidal curvature is suppressed in the core, as in the GR analysis of Danarianto \textit{et al.}\cite{ILoveCurvature2025}.

The Ricci scalar is the most EOS-sensitive invariant because $\mathcal{R}=\kappa(\mathcal{E}_{\mathrm{eff}}-3P_{\mathrm{eff}})$ encodes the effective trace of $T_{\mu\nu}^{\mathrm{eff}}$\cite{TraceCurvature}. In GR, $\mathcal{R}=24\pi\mathcal{E}\Delta$ and therefore tracks the fluid trace anomaly; in EMSG it tracks the \textit{effective} pressure--energy imbalance that sources the field equations. Softer models such as IOPB-I remain predominantly positive, whereas stiff EOSs such as NL3 develop large negative central values, $\mathcal{R} \sim -1.5 \times 10^{-8}\,\mathrm{m}^{-2}$, consistent with the Garibay \textit{et al.}\cite{TraceCurvature} result that about $50\%$ of viable EOSs admit negative Ricci curvature somewhere in the interior. Negative $\mathcal{R}$ marks a region where $P_{\mathrm{eff}}/\mathcal{E}_{\mathrm{eff}}>1/3$, i.e., where the geometry is sourced as if the effective fluid were stiffer than the conformal point. Because the sign change follows $\mathcal{E}_{\mathrm{eff}}-3P_{\mathrm{eff}}$ rather than the fluid-sector trace $\mathcal{E}-3P$ of Eq.~\eqref{eq:Delta}, a given fluid state can source negative effective Ricci curvature even when $\Delta$ is only mildly negative, or vice versa---a direct illustration of the matter--geometry separation of Sec.~\ref{sec:matter-geometry}. Overall, the hierarchy $\mathcal{K}>\mathcal{F}>\mathcal{W}\sim\mathcal{R}$ persists across the EOS set. The five hadronic RMF models used here are smooth and contain no explicit deconfinement transition; additional structure would appear for hybrid EOSs as shown by Ghosh \textit{et al.}\cite{Ghosh2026,HQPT,Anglani}.

\subsection{Radial profiles of curvature invariants}
\label{sec:curv-radial}
\begin{figure*}
    \centering
    \includegraphics[width=\textwidth]{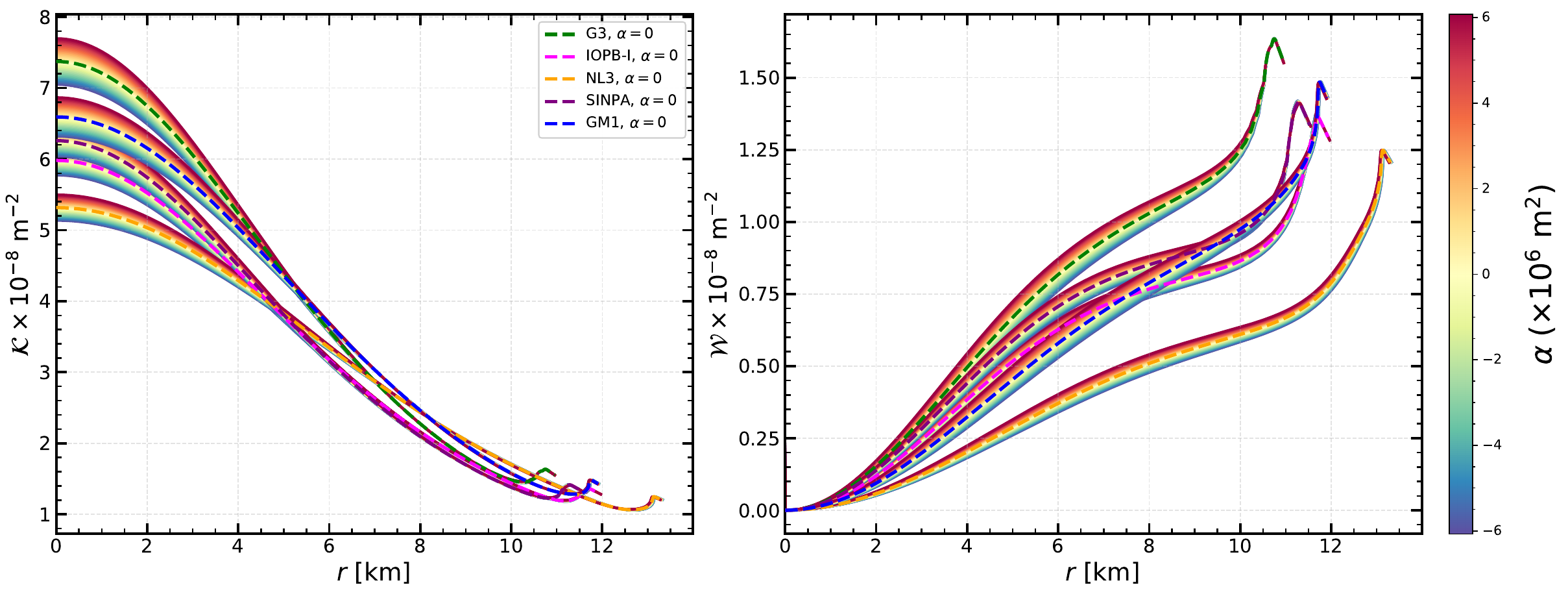}
    \caption{Radial profiles of the Kretschmann scalar $\mathcal{K}$ (left) and Weyl scalar $\mathcal{W}$ (right) in EMSG neutron stars.
    Each panel shows $\mathcal{I}(r)$ versus radius $r$ in km for five EOSs.
    Colored curves span $-6.07\times10^{6}\,\mathrm{m}^{2}\leq\alpha\leq 6.07\times10^{6}\,\mathrm{m}^{2}$; dashed curves denote $\alpha=0$.
    Curvature invariants are computed from $(\mathcal{E}_{\mathrm{eff}},P_{\mathrm{eff}})$ in units of $10^{-8}\,\mathrm{m}^{-2}$.}
    \label{radKW}
\end{figure*}

\begin{figure*}
    \centering
    \includegraphics[width=\textwidth]{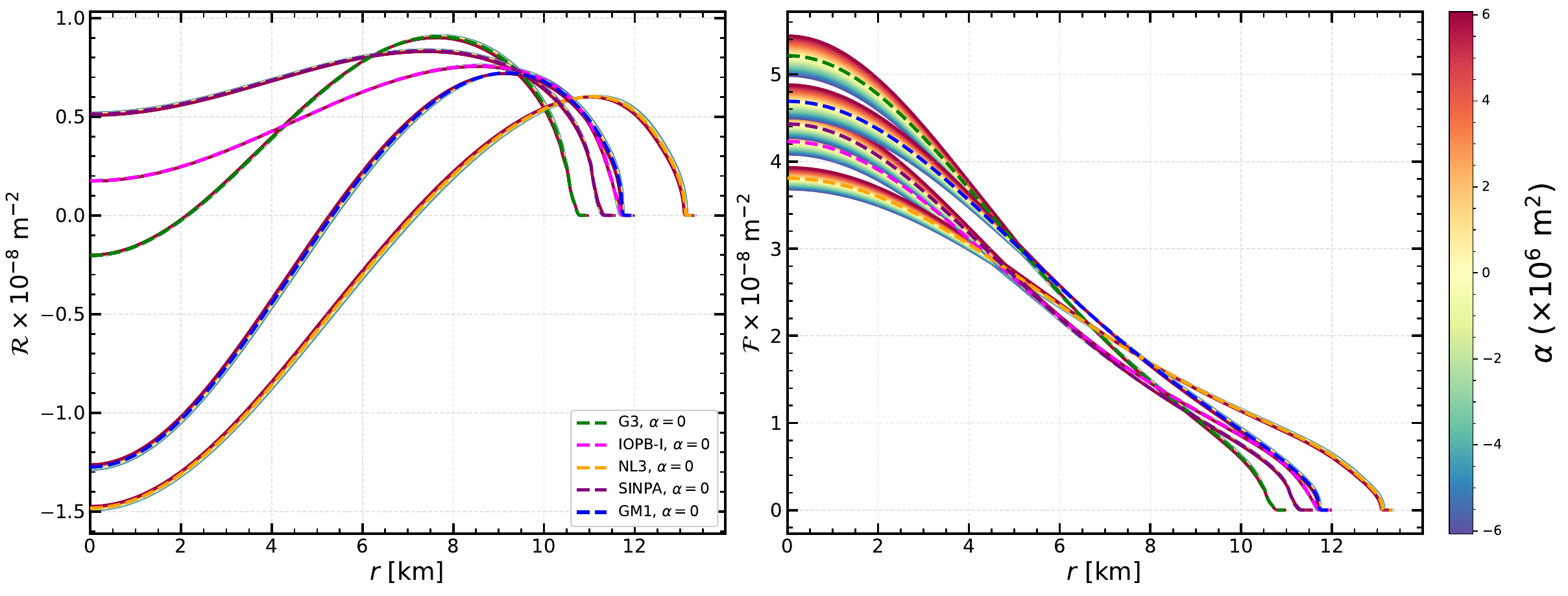}
    \caption{Radial profiles of the Ricci scalar $\mathcal{R}$ (left) and Ricci contraction $\mathcal{F}$ (right) in EMSG neutron stars.
    Curve organization follows Fig.~\ref{radKW}: radius $r$ in km on the horizontal axis and units of $10^{-8}\,\mathrm{m}^{-2}$ for both invariants.}
    \label{radRF}
\end{figure*}

Figures~\ref{radKW} and~\ref{radRF} display the radial profiles of the four invariants for each EOS, with colored curves spanning the coupling range of Eq.~\eqref{eq:alpha-range} and dashed curves showing the GR limit $\alpha=0$. Danarianto \textit{et al.}\cite{ILoveCurvature2025} reported in GR that $\mathcal{K}$ peaks at the center while $\mathcal{W}$ peaks in the outer layers, and that volume-averaged curvature scalars correlate strongly with $\bar{I}$ and $\Lambda$. The EMSG profiles preserve this qualitative ordering but acquire an $\alpha$-dependent shift because the metric responds to $T_{\mu\nu}^{\mathrm{eff}}$.
The Kretschmann scalar is maximal at the center and decreases monotonically outward, with $\mathcal{K}_{c}$ ranging from $\sim 5 \times 10^{-8}\,\mathrm{m}^{-2}$ to $\sim 7.5 \times 10^{-8}\,\mathrm{m}^{-2}$ depending on the EOS and on $\alpha$. This falloff follows from Eq.~\eqref{eq:Kretschmann}: the matter terms $\propto(\mathcal{E}_{\mathrm{eff}}^2+P_{\mathrm{eff}}^2)$ peak where the effective energy density peaks, while the exterior approaches the vacuum limit $\mathcal{K}_{\mathrm{surf}}^2=48M^2/r^6$\cite{ILoveCurvature2025}. Positive $\alpha$ uniformly enhances central $\mathcal{K}$, whereas negative $\alpha$ suppresses it; the separation between colored EMSG curves and the dashed $\alpha=0$ baselines is largest for NL3 and GM1, as expected from the $\mathcal{O}(\alpha\mathcal{E}^2)$ structure of the effective source\cite{Ghosh2026,EMSG_NAlam,Pinku-PRD_2023}.

The Weyl scalar shows the opposite radial trend. It vanishes at the center because $m(r)/r^3\to 4\pi\mathcal{E}_{\mathrm{eff}}/3$ as $r\to 0$ for regular spherical solutions\cite{Curvature3,ILoveCurvature2025}, then peaks near the surface at $r/R \gtrsim 0.95$ with $\mathcal{W}_{\max} \sim (1.2\text{--}1.6)\times 10^{-8}\,\mathrm{m}^{-2}$. This outer peak marks the region that dominantly controls the tidal Love number through the I--Love--$\mathcal{Q}$ relations\cite{Yagi:2013bca,Kent_yagi_2013,Harish-fmode_2022,Kokkotas1999}. EMSG preserves the GR picture---tidal curvature in the envelope, Ricci-dominated core---but shifts the amplitude of both invariants through the modified effective variables, so a change in $\alpha$ alters the outer-layer geometry that enters $\Lambda$ through Eqs.~\eqref{eq:tidal-pert}--\eqref{eq:Lambda}.

The Ricci scalar is strongly EOS dependent and can change sign within the interior for stiff models, signaling $P_{\mathrm{eff}}/\mathcal{E}_{\mathrm{eff}}>1/3$\cite{TraceCurvature,PhysRevLett.129.252702}. The radial location of the sign change differs between EOSs because the nuclear symmetry-energy slope shifts the balance between $\mathcal{E}_{\mathrm{eff}}$ and $P_{\mathrm{eff}}$\cite{NL3EOS,IOPB-IEOS}. The Ricci contraction $\mathcal{F}$ remains positive definite and decreases outward, with central values $\mathcal{F}_{c} \sim (3.8\text{--}5.5)\times 10^{-8}\,\mathrm{m}^{-2}$. Because $\mathcal{W}$ depends on the difference between $m(r)/r^3$ and $\mathcal{E}_{\mathrm{eff}}$, modified gravity can redistribute the radial location of the tidal peak even when the central $\mathcal{K}$ change is modest; this helps explain why the $\Lambda$- and $\bar{I}$-selected sequences in Sec.~\ref{sec:TA-observables} respond differently to $\alpha$. All four invariants vary smoothly with $\alpha$ over the interval of Eq.~\eqref{eq:alpha-range}, indicating that the EMSG solutions remain regular throughout the explored range\cite{Ghosh2026,EMSG_NAlam}.

\subsection{Pressure-to-energy ratio and trace anomaly profiles}
\label{sec:TA-profiles}
\begin{figure*}
    \centering
    \includegraphics[width=\textwidth]{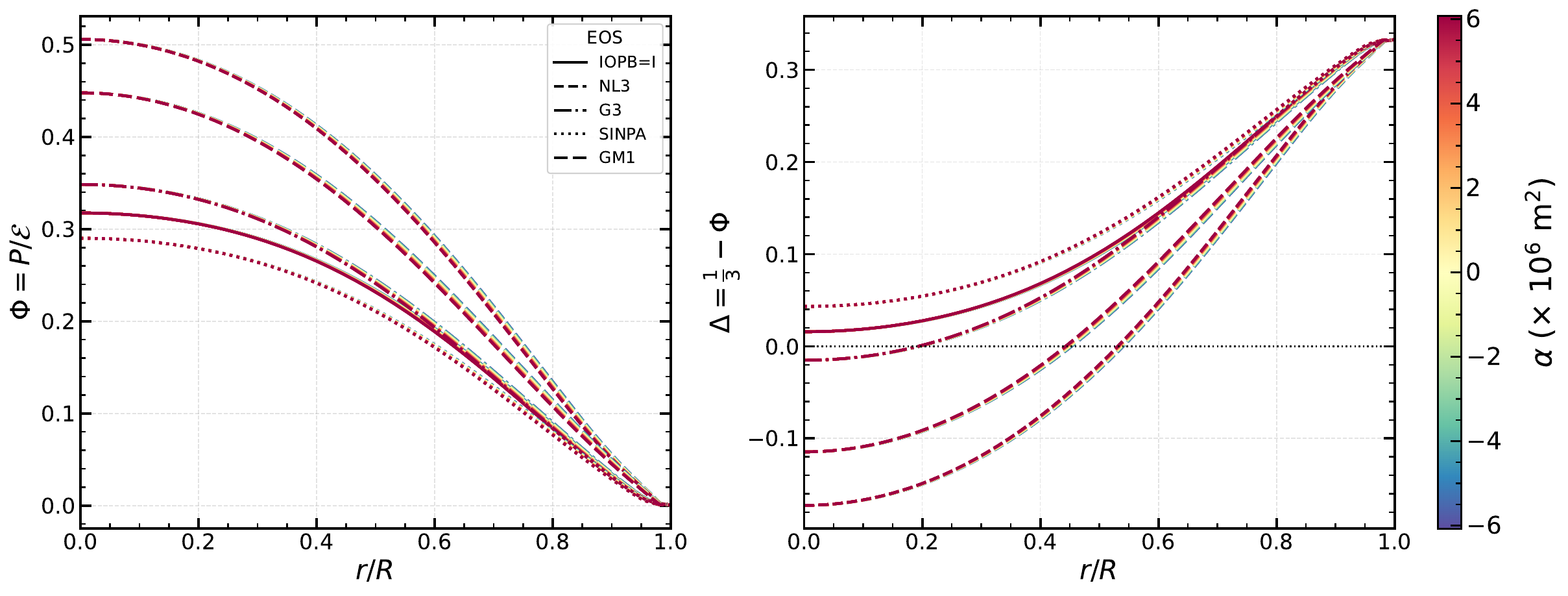}
    \caption{Radial profiles of the pressure-to-energy ratio $\phi=P/\mathcal{E}$ (left) and trace anomaly $\Delta=1/3-\phi$ (right) for EMSG neutron stars.
    Results are shown for NL3, IOPB-I, G3, SINPA, and GM1 as functions of $r/R$; colored curves span $-6.07\times10^{6}\,\mathrm{m}^{2}\leq\alpha\leq 6.07\times10^{6}\,\mathrm{m}^{2}$.
    The trace anomaly is computed exclusively from the fluid variables $(\mathcal{E},P)$; dashed curves indicate $\alpha=0$.}
    \label{TArad}
\end{figure*}

We now turn from the geometry sector to the fluid-sector trace anomaly. Figure~\ref{TArad} shows the radial evolution of $\phi(r)=P(r)/\mathcal{E}(r)$ and $\Delta(r)=1/3-\phi(r)$ for each EOS at representative values of $\alpha$. Cai \textit{et al.}\cite{Cai_2026,Cai2024compactness} proved in GR that $\phi$ must decrease outward from the center for any viable static solution, so $\Delta$ attains a minimum in the core; Ren and Lin\cite{ren2026tracingtraceanomalydense} showed that the full profile is nevertheless organized by global observables to within about $10\%$ EOS scatter. The question here is whether the modified TOV system preserves that GR morphology numerically while introducing an $\alpha$-dependent redistribution of $(\mathcal{E},P)$.
For all EOSs and every accepted $(\mathrm{EOS},\alpha)$ solution in our scan, $\phi=P/\mathcal{E}$ decreases monotonically from the center toward the surface and $\Delta$ increases outward, reproducing the GR trend established by Cai \textit{et al.}\cite{Cai_2026,Cai2024compactness} but without invoking their GR-only proof in modified gravity. Because $\Delta$ is evaluated from the unmodified fluid variables, the EMSG coupling does not alter its definition; any $\alpha$ dependence enters only through the modified hydrostatic equilibrium that redistributes $(\mathcal{E},P)$. Positive $\alpha$ tends to deepen the central gravitational potential and compress the core, increasing $\phi_c$ and driving $\Delta_c$ more negative for stiff EOSs, whereas negative $\alpha$ produces the opposite trend\cite{EMSG_NAlam,Ghosh2025,Pinku-PRD_2023}.

The central values are strongly EOS dependent. NL3 reaches the stiffest core with $\phi_{c}\gtrsim 0.50$ and $\Delta_{c}\lesssim -0.15$; GM1 follows with $\phi_{c}\sim 0.45$ and $\Delta_{c}\sim -0.10\text{--}0.12$; G3 is intermediate; whereas softer EOSs such as IOPB-I and SINPA remain near $\phi_{c}\sim 0.28\text{--}0.35$ with $\Delta_{c}\sim +0.02\text{--}0.05$\cite{PhysRevLett.129.252702,TraceCurvature}. Fujimoto \textit{et al.}\cite{PhysRevLett.129.252702} conjectured that $\Delta$ may remain nonnegative in NS matter, but subsequent ensemble studies by Garibay \textit{et al.}\cite{TraceCurvature} and Ren and Lin\cite{ren2026tracingtraceanomalydense} showed that stiff EOSs can produce negative $\Delta_c$, with the central value passing through zero near zero-curvature compactness $\mathscr{C}_0=0.26911\pm0.00004$\cite{TraceCurvature} and $C\simeq0.26\text{--}0.28$\cite{ren2026tracingtraceanomalydense,Cai_2026}. NL3 and GM1 are also the most sensitive to $\alpha$ because their large central $\mathcal{E}$ maximizes the $\alpha\mathcal{E}^2$ source term\cite{Ghosh2025,EMSG_NAlam,Bharat_and_Landry}. Near the surface, $\phi\to 0$ and $\Delta\to 1/3$ for every EOS, confirming that the outer envelope approaches the dilute limit independently of core stiffness and modified-gravity coupling.

\subsection{Dependence on compactness, tidal deformability, and moment of inertia}
\label{sec:TA-observables}
\begin{figure*}
    \centering
    \includegraphics[width=\textwidth]{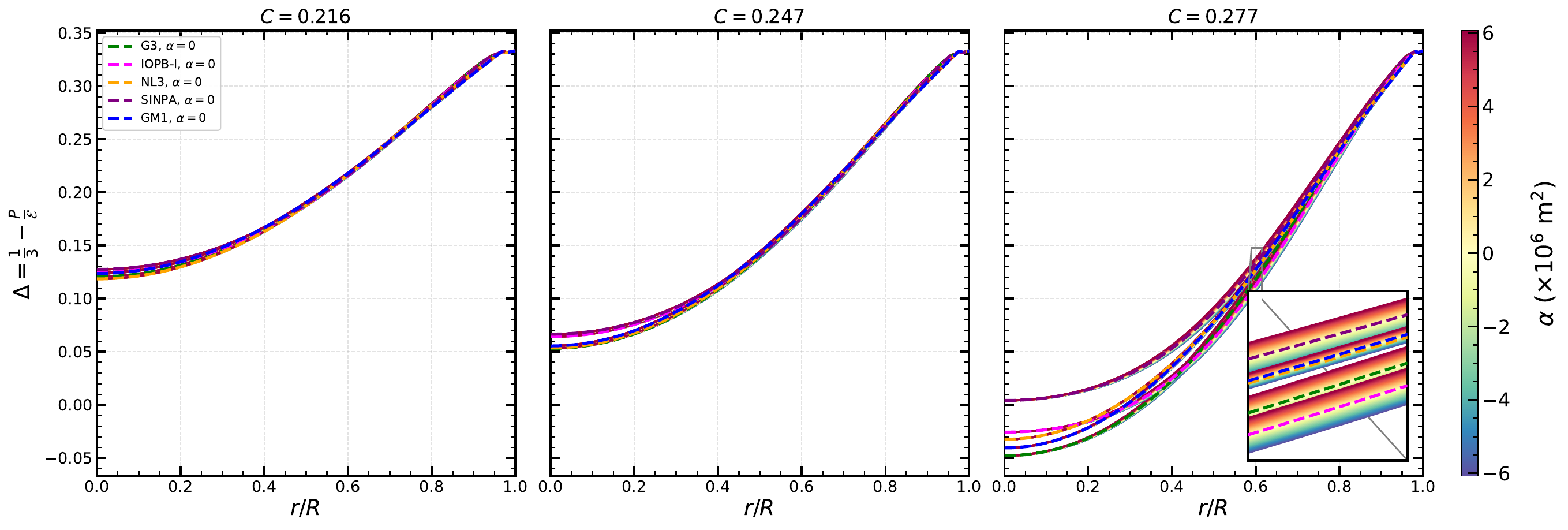}
    \caption{Trace anomaly profiles $\Delta(r/R)$ for neutron-star sequences with fixed compactness $C=0.216$, $0.247$, and $0.277$ (left to right).
    Each panel shows five EOSs; colored curves span $-6.07\times10^{6}\,\mathrm{m}^{2}\leq\alpha\leq 6.07\times10^{6}\,\mathrm{m}^{2}$, and dashed curves indicate the GR baseline $\alpha=0$.
    Sequence targets follow the construction of Ren and Lin\cite{ren2026tracingtraceanomalydense}; $\Delta$ is defined from $(\mathcal{E},P)$ only.}
    \label{TAwithCompactness}
\end{figure*}

\begin{figure*}
    \centering
    \includegraphics[width=\textwidth]{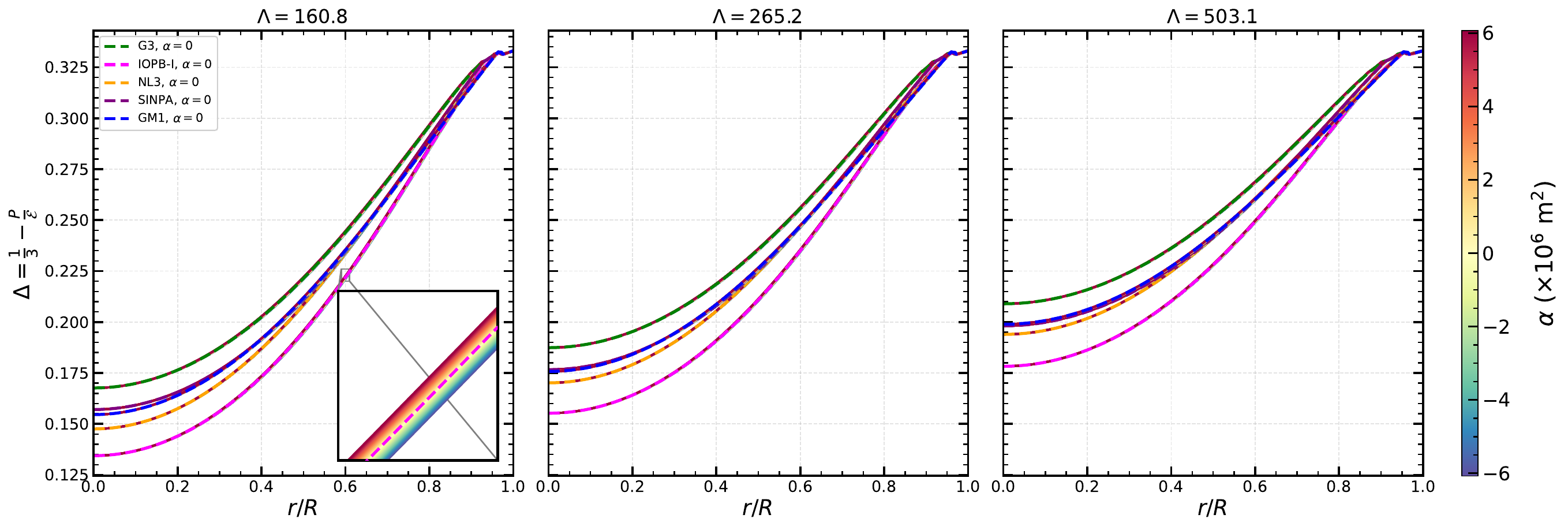}
    \caption{Trace anomaly profiles $\Delta(r/R)$ for sequences with fixed tidal deformability $\Lambda=160.8$, $265.2$, and $503.1$ (left to right), bracketing the GW170817 range for a $1.4\,M_\odot$ star\cite{PhysRevLett.121.091102,GW170817}.
    Curve coding follows Fig.~\ref{TAwithCompactness}.}
    \label{TAwithTidalDeformability}
\end{figure*}

\begin{figure*}
    \centering
    \includegraphics[width=\textwidth]{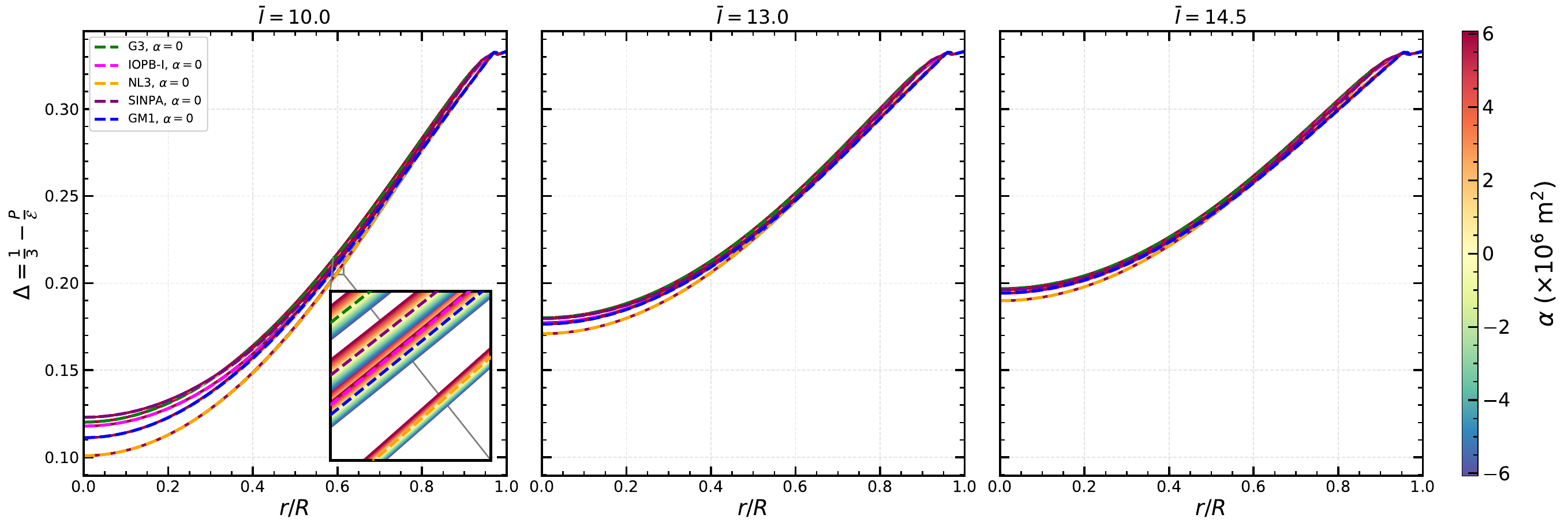}
    \caption{Trace anomaly profiles $\Delta(r/R)$ for sequences with fixed normalized moment of inertia $\bar{I}=10$, $13$, and $14.5$ (left to right).
    Curve coding is identical to Figs.~\ref{TAwithCompactness} and~\ref{TAwithTidalDeformability}.}
    \label{TAwithMomentofInertia}
\end{figure*}

Ren and Lin\cite{ren2026tracingtraceanomalydense} demonstrated that fixing any one of $C$, $\Lambda$, or $\bar{I}$ organizes the full radial profile $X(r/R)=P/\mathcal{E}$---and hence $\Delta(r)$---to within about $10\%$ across 45 observationally viable GR EOSs. Here we repeat that exercise in EMSG: for each target observable we map to a central pressure using GR reference integrations (Sec.~\ref{sec:tidal}), then solve the modified TOV system at each $(\mathrm{EOS},\alpha)$ and extract $\Delta(r)$ from the fluid variables alone. The EMSG coupling therefore enters only through the reshaped hydrostatic equilibrium, not through a redefinition of $\Delta$. Figures~\ref{TAwithCompactness}--\ref{TAwithMomentofInertia} present the resulting profile bands; positive $\alpha$ generally increases $\Delta(r/R)$ relative to the GR baseline (dashed curves), whereas negative $\alpha$ decreases it, with the largest departures appearing in the intermediate region $0.3\lesssim r/R \lesssim 0.8$.

\subsubsection{Fixed compactness}
\label{sec:TA-C}
We begin with sequences constructed at fixed compactness, following the procedure of Ren and Lin\cite{ren2026tracingtraceanomalydense}. Figure~\ref{TAwithCompactness} shows that, at the lowest compactness considered, $C=0.216$, the $\Delta(r/R)$ profiles remain positive and exhibit only mild EOS scatter, closely resembling the GR quasi-universal surfaces ($\sigma=7.62\times10^{-3}$ for the $(C,z,X)$ correlation in Ref.~\cite{ren2026tracingtraceanomalydense}). The EMSG curves remain near the dashed $\alpha=0$ baselines because the central $\mathcal{E}$ is too small for the $\alpha\mathcal{E}^2$ correction to compete with the linear fluid source. As $C$ increases to $0.247$ and $0.277$, the central trace anomaly moves toward negative values for several EOSs, mirroring the GR result that $\Delta_c$ becomes negative for $C\simeq0.26\text{--}0.28$\cite{ren2026tracingtraceanomalydense,Cai_2026} and passes through zero near $\mathscr{C}_0=0.26911\pm0.00004$\cite{TraceCurvature}, while the $\alpha$ sequences fan out in the intermediate radial region because the effective source scales with $\mathcal{E}^2$ in the dense core\cite{EMSG_NAlam,Ghosh2026}.

\subsubsection{Fixed tidal deformability}
\label{sec:TA-Lambda}
Turning next to fixed tidal deformability, Fig.~\ref{TAwithTidalDeformability} displays sequences with $\Lambda=160.8$, $265.2$, and $503.1$, bracketing the GW170817 inference for a $1.4\,M_{\odot}$ star\cite{PhysRevLett.121.091102,GW170817,Annala_2020,PhysRevX.9.011001}. Because $\Lambda$ is dominated by the outer tidal layer\cite{Yagi:2013bca,Kent_yagi_2013}, Ren and Lin\cite{ren2026tracingtraceanomalydense} found that fixing $\ln\Lambda$ nevertheless organizes the full $\Delta(r)$ profile to within about $10\%$ EOS scatter ($\sigma=6.38\times10^{-3}$ for the $(\Lambda,z,X)$ correlation). At the smallest value, $\Lambda\simeq160$, the $\alpha$ fanning is strongest; as $\Lambda$ increases toward $\sim503$, the curves converge and the EMSG splitting weakens, reproducing the quasi-universal GR grouping. The outer layers at $r/R\gtrsim 0.8$ remain comparatively EOS independent, consistent with the surface boundary condition $X(u=0)=0$ imposed in the fitting surfaces of Ref.~\cite{ren2026tracingtraceanomalydense}.

\subsubsection{Fixed normalized moment of inertia}
\label{sec:TA-Ibar}
Finally, Fig.~\ref{TAwithMomentofInertia} shows sequences at fixed $\bar{I}=10$, $13$, and $14.5$. Because $\bar{I}$ weights the entire interior mass profile and is only weakly correlated with $C$ and $\Lambda$ in the I--Love--$\mathcal{Q}$ framework\cite{Yagi:2013bca,Kent_yagi_2013,Kent_yagi_2015,Haskell_2015,ILoveCurvature2025,Gupta_2018}, these panels provide a complementary probe of the core. Ren and Lin\cite{ren2026tracingtraceanomalydense} obtain the tightest profile organization at fixed $\ln\bar{I}$ ($\sigma=6.02\times10^{-3}$). In EMSG, the $\bar{I}=10$ panel shows the largest $\alpha$ splitting and EOS separation, whereas the $\bar{I}=14.5$ profiles become nearly degenerate except in the inner core. Taken together, Figs.~\ref{TAwithCompactness}--\ref{TAwithMomentofInertia} show that EMSG does not qualitatively alter the thermodynamic organization of $\Delta$ established in GR; it introduces controlled, $\alpha$-dependent deformations of intrinsically stable profile bands\cite{Ghosh2025,EMSG_NAlam}.

\subsection{Curvature--trace anomaly correspondence}
\label{sec:curv-delta}
\begin{figure*}
    \centering
    \includegraphics[width=\textwidth]{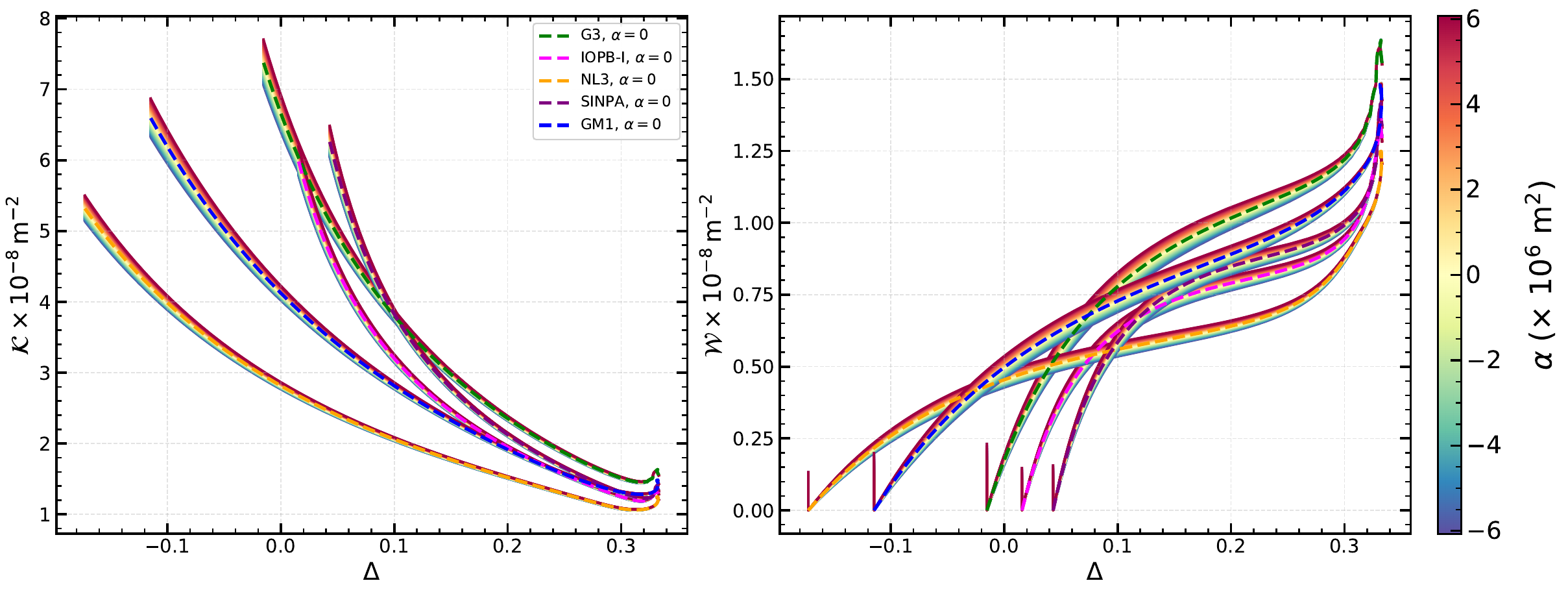}
    \caption{Kretschmann scalar $\mathcal{K}$ (left) and Weyl scalar $\mathcal{W}$ (right) plotted against the trace anomaly $\Delta$.
    Colored curves span the EMSG coupling range $-6.07\times10^{6}\,\mathrm{m}^{2}\leq\alpha\leq 6.07\times10^{6}\,\mathrm{m}^{2}$; dashed curves show the GR limit $\alpha=0$ for each EOS.
    Curvature invariants are computed from $(\mathcal{E}_{\mathrm{eff}},P_{\mathrm{eff}})$ and scaled to units of $10^{-8}\,\mathrm{m}^{-2}$.}
    \label{KWDelta}
\end{figure*}

\begin{figure*}
    \centering
    \includegraphics[width=\textwidth]{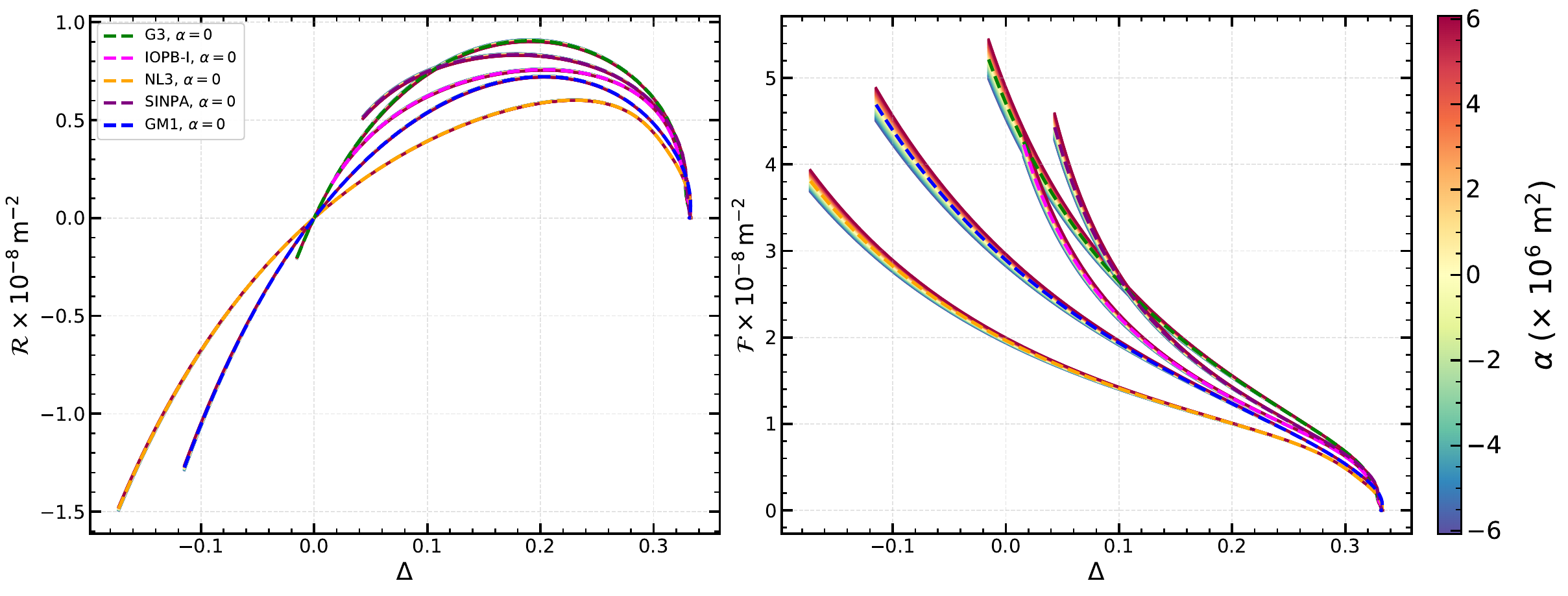}
    \caption{Ricci scalar $\mathcal{R}$ (left) and Ricci contraction $\mathcal{F}$ (right) as functions of the trace anomaly $\Delta$.
    Curve organization follows Fig.~\ref{KWDelta}; units are $10^{-8}\,\mathrm{m}^{-2}$.}
    \label{RFDelta}
\end{figure*}

Garibay \textit{et al.}\cite{TraceCurvature} showed in GR that curvature invariants correlate systematically with the trace anomaly when both are built from the same $(\mathcal{E},P)$. In EMSG the metric responds to $(\mathcal{E}_{\mathrm{eff}},P_{\mathrm{eff}})$ while we keep $\Delta$ on the fluid sector. If the GR organization survives, radial sequences with different $(\alpha,\mathrm{EOS})$ should collapse onto organized bands when $\mathcal{I}$ is plotted against $\Delta$ rather than against $r$ or $\rho_B$. Figures~\ref{KWDelta} and~\ref{RFDelta} present this test for all four invariants.
The Kretschmann scalar decreases monotonically with increasing $\Delta$: in the stiffest cores ($\Delta \lesssim -0.1$) one finds $\mathcal{K} \sim (5\text{--}7)\times 10^{-8}\,\mathrm{m}^{-2}$, whereas near the surface, where $\Delta \to 1/3$, $\mathcal{K}$ falls to $\sim 1.2 \times 10^{-8}\,\mathrm{m}^{-2}$. Plotting $\mathcal{I}$ against $\Delta$ replaces the radial coordinate with a local thermodynamic label\cite{ren2026tracingtraceanomalydense,TraceCurvature}. The monotonic $\mathcal{K}(\Delta)$ trend parallels the GR organization of Garibay \textit{et al.}\cite{TraceCurvature}, although the EMSG curves are vertically shifted by $\alpha$ because the invariants are built from $(\mathcal{E}_{\mathrm{eff}},P_{\mathrm{eff}})$. Positive $\alpha$ raises the entire band; negative $\alpha$ lowers it. Keeping $\Delta$ on the horizontal axis---rather than an effective trace built from $(\mathcal{E}_{\mathrm{eff}},P_{\mathrm{eff}})$---is what makes the EMSG result nontrivial and directly comparable to the GR literature.

The Weyl scalar increases with $\Delta$ and develops a branching structure at intermediate values, marking the transition from the Ricci-dominated core to the tidal-dominated crust where $\mathcal{W}\propto[(6m/r^3)-8\pi\mathcal{E}_{\mathrm{eff}}]^2$ becomes large\cite{Curvature3,ILoveCurvature2025}. The Ricci scalar is strongly non-monotonic: different EOSs produce distinct branches at negative $\Delta$ because $\mathcal{R}$ tracks the \textit{effective} trace while $\Delta$ is built from $(\mathcal{E},P)$. This branch structure is a direct signature of the matter--geometry separation. By contrast, the Ricci contraction $\mathcal{F}$ shows the tightest $\mathcal{I}(\Delta)$ organization and the smallest EOS scatter in our EMSG scan, as in the GR $\mathcal{R}$--$\Delta$ analysis of Garibay \textit{et al.}\cite{TraceCurvature}, where the related invariant $\mathcal{J}$ also tracks $\Delta$ more smoothly than $\mathcal{R}$.

Taken together, Figs.~\ref{KWDelta} and~\ref{RFDelta} show that radial sequences collapse onto organized single-parameter bands when expressed as $\mathcal{I}(\Delta)$. Because $\Delta$ is constructed only from $(\mathcal{E},P)$ while $\mathcal{I}$ depends on $(\mathcal{E}_{\mathrm{eff}},P_{\mathrm{eff}})$, this organization constitutes a nontrivial extension of the GR correspondence of Garibay \textit{et al.}\cite{TraceCurvature} into EMSG. The relations are not strictly universal---Garibay \textit{et al.}\cite{TraceCurvature} themselves report EOS scatter in the $\mathcal{R}$--$\Delta$ plane, and EMSG adds an $\alpha$-dependent vertical shift---but the residual dependence remains moderate over most of the explored range. Visually, $\mathcal{F}(\Delta)$ is the tightest band and $\mathcal{R}(\Delta)$ the most EOS sensitive, as in the GR $\mathcal{J}(\Delta)$ versus $\mathcal{R}(\Delta)$ comparison of Garibay \textit{et al.}\cite{TraceCurvature}. The looser $\mathcal{R}(\Delta)$ branches indicate where EOS-sensitive core physics or phase transitions\cite{HQPT,alford2008color} are most likely to distinguish GR from EMSG despite similar global masses and radii.

\subsection{Observational sequences and ultracompact limits}
\label{sec:obs-sequences}
\begin{figure*}
    \centering
    \includegraphics[width=\textwidth]{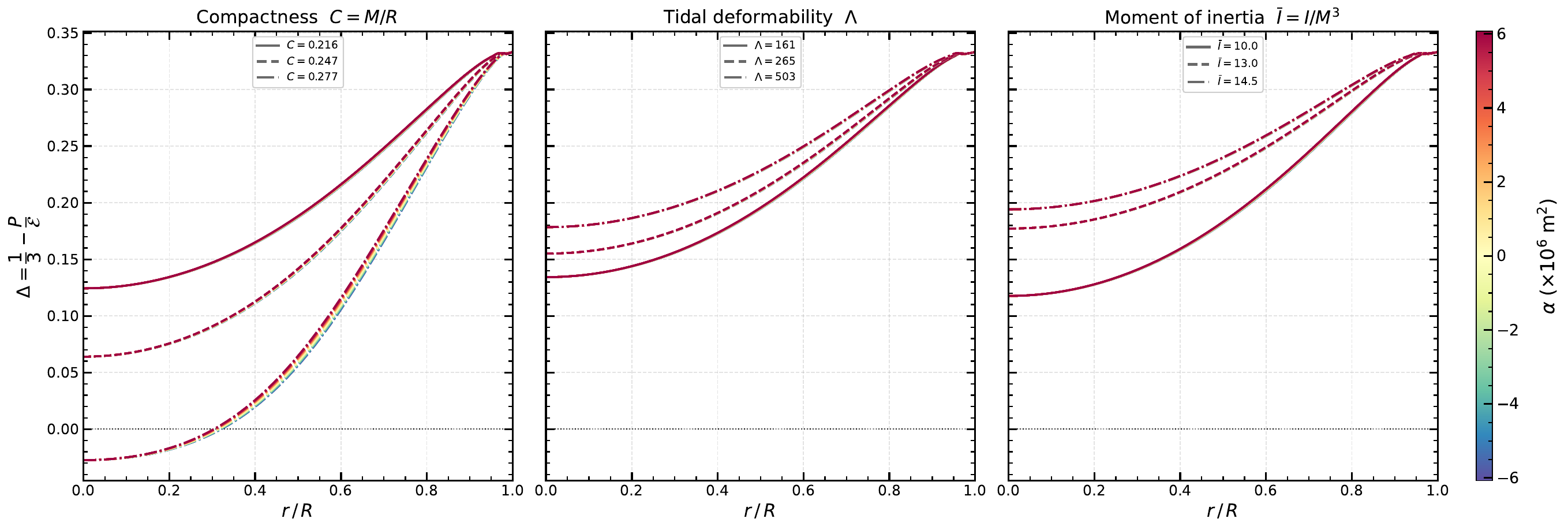}
    \caption{Combined trace anomaly profiles $\Delta(r/R)$ for realistic EMSG neutron-star sequences.
    Rows correspond to fixed compactness $C$ (top), tidal deformability $\Lambda$ (middle), and normalized moment of inertia $\bar{I}$ (bottom); columns show $C=0.216$, $0.247$, and $0.277$, $\Lambda=160.8$, $265.2$, and $503.1$, and $\bar{I}=10$, $13$, and $14.5$, respectively, as labeled in each panel.
    Five EOSs are shown in each panel with the same $\alpha$-color coding as Figs.~\ref{TAwithCompactness}--\ref{TAwithMomentofInertia}.}
    \label{RealisticCTM}
\end{figure*}

\begin{figure*}
    \centering
    \includegraphics[width=\textwidth]{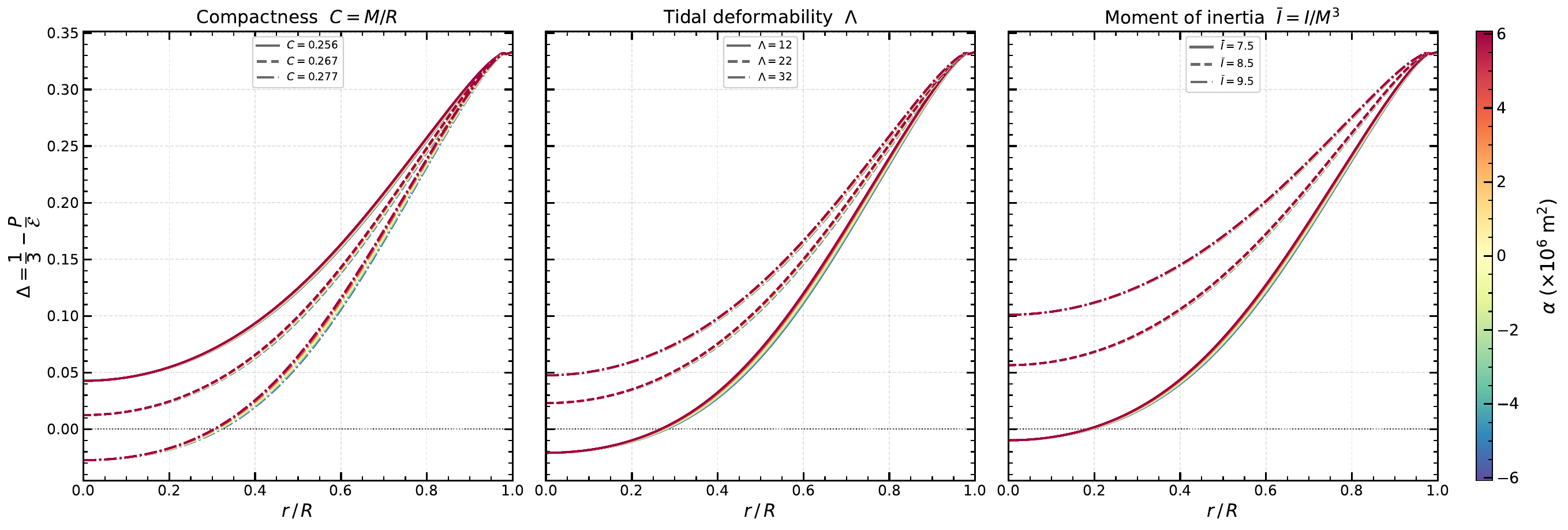}
    \caption{Trace anomaly profiles $\Delta(r/R)$ for ultracompact theoretical sequences, where EMSG-induced profile splitting is largest.
    Row and column organization follows Fig.~\ref{RealisticCTM}, with targets $C=0.256$, $0.267$, and $0.277$ (top row), $\Lambda=12$, $22$, and $32$ (middle row), and $\bar{I}=7.5$, $8.5$, and $9.5$ (bottom row), as labeled in each panel.}
    \label{TheoreticalCTM}
\end{figure*}

Figures~\ref{RealisticCTM} and~\ref{TheoreticalCTM} combine the compactness, tidal, and moment-of-inertia sequences in a format suited to comparison with the observational applications of Ren and Lin\cite{ren2026tracingtraceanomalydense}. The realistic panels use the same $(C,\Lambda,\bar{I})$ targets as Figs.~\ref{TAwithCompactness}--\ref{TAwithMomentofInertia}; the theoretical panels use more compact sequences with $(C,\Lambda,\bar{I})=(0.256,0.267,0.277)$, $(12,22,32)$, and $(7.5,8.5,9.5)$, where EMSG profile splitting is strongest.
In the realistic panels of Fig.~\ref{RealisticCTM}, increasing $C$ drives the profiles toward more negative central $\Delta$, consistent with Fig.~\ref{TAwithCompactness} and with the GR analyses of Cai \textit{et al.}\cite{Cai_2026,Cai2024compactness} and Ren and Lin\cite{ren2026tracingtraceanomalydense}. Ren and Lin applied their quasi-universal surfaces to NICER targets PSR~J0030+0451 and PSR~J0740+6620 and inferred $\Delta_c=0.1770^{+0.0365}_{-0.0432}$ for a $1.4\,M_\odot$ star from the GW170817 tidal constraint. Our EMSG sequences at comparable $\Lambda$ remain consistent with this range for soft-to-intermediate EOSs, while stiff models with negative $\Delta_c$ probe the opposite side of the conformal boundary. NICER radii and GW $\Lambda$ measurements therefore constrain $\Delta_c$ indirectly in GR; Fig.~\ref{RealisticCTM} shows how EMSG widens the allowable band through $\alpha$ without changing the monotonic radial shape\cite{Miller_2019,Riley_2021,Miller_2021,PhysRevLett.121.091102,Annala_2020}.

In the ultracompact regime of Fig.~\ref{TheoreticalCTM}, the colored $\alpha$ curves separate strongly from the dashed GR baselines and the central $\Delta$ values become increasingly negative for stiff EOSs, as expected from the $\mathcal{O}(\alpha\mathcal{E}^2)$ scaling in Eqs.~\eqref{eq:Eeff}--\eqref{eq:Peff}\cite{Ghosh2026,EMSG_NAlam,Pinku-PRD_2023}. This is the same region where Garibay \textit{et al.}\cite{TraceCurvature} found the strongest $\mathcal{R}$--$\Delta$ correlations, with $\Delta_{\min}=-0.227$ in their ensemble, and where Ren and Lin\cite{ren2026tracingtraceanomalydense} report $\Delta_c<0$ for highly compact configurations. Modified-gravity signatures in $\Delta(r)$ therefore concentrate in stiff, compact stars rather than being uniformly distributed across the mass--radius plane. Even here, however, the radial ordering of the profiles is unchanged and no irregular features appear. Existing bounds on $\alpha$ from cosmology and binary pulsars\cite{EMSG_OAkarsu,Nazari1,Board2017} remain far tighter than what the present profile splitting alone would require; the ultracompact panels should therefore be read as illustrating the \textit{direction} of the EMSG effect rather than as an independent constraint on $\alpha$.

\section{Scope and Limitations}
\label{sec:limitations}

Several simplifying assumptions bound the scope of the present analysis and should be kept in mind when interpreting the results.

\subsubsection{Equation-of-state set}
We employ five hadronic relativistic mean-field EOSs without explicit phase transitions, hyperonic degrees of freedom, or quark matter. The $\mathcal{R}(\Delta)$ branch structure and the magnitude of ultracompact profile splitting are therefore EOS dependent, and extending the study to hybrid or quark-hadron models\cite{HQPT,alford2008color,Anglani} is a natural next step.

\subsubsection{Perturbative approximations}
All models are static, spherically symmetric, and nonrotating. Tidal deformability and normalized moment of inertia enter through the GR perturbation formalism of Sec.~\ref{sec:tidal}; the sequence targets $(C,\Lambda,\bar{I})$ are mapped to central conditions with GR reference integrations before the EMSG background profiles are evaluated. A fully self-consistent EMSG treatment of tidal and rotational perturbations would modify both the background and the response sector and could alter the quoted $\alpha$ splitting at fixed global observables.

\subsubsection{Coupling parameter and observational bounds}
The coupling interval of Eq.~\eqref{eq:alpha-range} is chosen to bracket prior EMSG neutron-star analyses\cite{Ghosh2026,EMSG_NAlam,Pinku-PRD_2023}. Cosmological and binary-pulsar constraints on $\alpha$ are typically far tighter than the profile splitting alone would require\cite{EMSG_OAkarsu,Nazari1,Board2017}. For sequences matched to GW170817 tidal deformabilities, the maximum $\Delta(r/R)$ separation between EMSG and GR profiles at fixed $\Lambda$ remains modest for soft-to-intermediate EOSs, whereas stiff ultracompact models show the largest deviations. The present work therefore illustrates how modified gravity \textit{would} deform thermodynamic--geometric relations if $\alpha$ were astrophysically relevant, rather than deriving a new independent bound on $\alpha$.

\subsubsection{Universality assessment}
The $\mathcal{I}(\Delta)$ organization reported in Sec.~\ref{sec:curv-delta} is assessed visually in Figs.~\ref{KWDelta} and~\ref{RFDelta}. A systematic scatter analysis across a larger EOS ensemble, along the lines of Garibay \textit{et al.}\cite{TraceCurvature} and Ren and Lin\cite{ren2026tracingtraceanomalydense}, remains for future work. Future studies could also combine NICER radii, gravitational-wave $\Lambda$ measurements, and pulsar timing constraints on $\bar{I}$ in a Bayesian framework to test whether the $\Delta$--curvature correspondence survives simultaneous EOS and modified-gravity inference\cite{Zhao_2022,Miller_2019}.

\section{Summary and Conclusion}
\label{SnC}
We have investigated whether the trace anomaly $\Delta$, defined only from the fluid sector, continues to organize interior spacetime geometry when neutron stars are modeled in energy-momentum squared gravity (EMSG). Adopting the matter--geometry separation of Sec.~\ref{sec:matter-geometry}, we computed $\Delta$ from the physical fluid variables and constructed the curvature invariants from the effective variables that source the modified TOV equations. For five relativistic mean-field equations of state over the EMSG coupling range studied in Sec.~\ref{sec:numerics}, three main conclusions emerge.

First, the fluid-sector trace-anomaly profiles increase monotonically from core to surface and remain equation-of-state structured in all accepted EMSG integrations, numerically mirroring the outward-increasing GR trend established model-independently by Cai \textit{et al.}\cite{Cai_2026,Cai2024compactness} and the quasi-universal profile organization of Ren and Lin\cite{ren2026tracingtraceanomalydense}, but they acquire a coupling-dependent splitting that grows with compactness and is strongest for stars with smaller tidal deformability and smaller normalized moment of inertia. Second, the curvature invariants evolve smoothly with baryon density and radius, preserve the GR radial ordering in which the Kretschmann scalar peaks in the core and the Weyl scalar is concentrated in the outer layers\cite{ILoveCurvature2025}, and respond continuously to the EMSG coupling throughout the explored range\cite{Ghosh2026,EMSG_NAlam}. Third, and most importantly, the invariants nevertheless collapse onto organized one-parameter bands when plotted against $\Delta$, extending the GR thermodynamic--geometric correspondence of Garibay \textit{et al.}\cite{TraceCurvature} into modified gravity. The Ricci contraction shows the tightest grouping; the Ricci scalar retains the largest equation-of-state sensitivity in the negative-$\Delta$ core, consistent with the $\mathcal{R}$--$\Delta$ scatter reported by Garibay \textit{et al.}\cite{TraceCurvature}.

Nonlinear gravity therefore shifts spacetime curvature without erasing the thermodynamic ordering already established in GR. The effect concentrates in stiff, ultracompact configurations, while observationally accessible sequences matched to NICER and GW170817 show only moderate coupling-induced deformation of the trace-anomaly profiles. Multimessenger constraints on mass, radius, and tidal deformability\cite{Miller_2019,Riley_2021,PhysRevLett.121.091102,Annala_2020,Jiang_2020}, together with existing EMSG bounds on $\alpha$ from cosmology and binary pulsars\cite{EMSG_OAkarsu,Nazari1,Board2017}, can refine this picture once a fully self-consistent EMSG perturbation treatment and a broader equation-of-state ensemble are in place (Sec.~\ref{sec:limitations}). Treating the trace anomaly and curvature invariants as complementary interior descriptors offers a practical framework for comparing GR and modified-gravity interpretations of future pulsar, X-ray, and gravitational-wave data\cite{PhysRevX.11.041050,GW190814,Tran_2022,Riahi_2019,Roupas2021,Fazlollahi2023,Mohanty_2024}.

\begin{acknowledgments}
The authors thank colleagues at NIT Rourkela for useful discussions on modified gravity and neutron star structure.
\end{acknowledgments}
\bibliographystyle{apsrev4-1}
\bibliography{main}

\end{document}